\documentclass[11pt,aps,english]{article}

\usepackage{graphicx,array}
\usepackage{mathrsfs} 
\usepackage{color}
\usepackage{latexsym}
\usepackage{amsthm}
\usepackage{amsmath}
\usepackage{amssymb}
\usepackage{dsfont}
\usepackage{epsfig}
\usepackage{subfig}
\usepackage{wrapfig}
\usepackage{slashed}
\usepackage{bbold}
\usepackage{psfrag}
\usepackage[svgnames]{xcolor}
\usepackage{xfrac}
\usepackage{multirow}
\usepackage{booktabs}
\usepackage[colorlinks=true,linkcolor=MediumBlue,citecolor=Green,urlcolor=violet]{hyperref}
\usepackage[normalem]{ulem}
\usepackage{footnote}
\usepackage{tikz-feynman}
\usepackage{xltabular}
\usepackage[utf8]{inputenc}
\usepackage{geometry}
\usepackage{tablefootnote}
\usepackage{float}
\usepackage{cite}
\usepackage{array}

\newcolumntype{P}[1]{>{\centering\arraybackslash}p{#1}}

\def\bea{\begin{eqnarray}}
\def\eea{\end{eqnarray}}
\def\beq{\begin{equation}}
\def\eeq{\end{equation}}
\def\be{\begin{equation}}
\def\ee{\end{equation}}

\def\mL{\mathcal{L}}
\def\mM{\mathcal{M}}

\def\mO{\mathcal{O}}

\def\1{\textbf{1}}
\def\2{\textbf{2}}
\def\3{\textbf{3}}
\def\4{\textbf{4}}
\def\5{\textbf{5}}
\def\6{\textbf{6}}
\def\7{\textbf{7}}
\def\8{\textbf{8}}
\def\9{\textbf{9}}

\newcommand{\red}[1]{{\color{red} #1}}



\begin{document}
\title{Top Yukawa Coupling at the Muon Collider}

\author{\vspace{1cm}Miranda Chen\thanks{mwqchen@ucdavis.edu} \and Da Liu\thanks{daeliu@ucdavis.edu}}
\date{\it{Center for Quantum Mathematics and Physics (QMAP)} \\ \it{University of California, Davis, CA 95616, USA}}

\maketitle
\begin{abstract}
We have presented a detailed study about the prospects for the measurement of the top Yukawa coupling in the vector boson fusion (VBF) production of a top quark pair at high energy muon colliders. By employing  the effective $W$ approximation and the high energy limit for the helicity amplitudes of the subprocess $W^+ W^- \rightarrow t \bar t$, we have derived the energy scaling of the  statistical signal significance in the presence of the anomalous couplings. The sensitivity on the top Yukawa coupling decreases as the bin energy increases. For the anomalous triple gauge boson couplings and the gauge-boson-fermion couplings  with $E^2$ energy growing behavior for the interference term, the signal significance has mild increase at the beginning and start to decrease for $\hat s_{t \bar t} \sim 0.2  \, s_{\mu^+ \mu^-} $. The 95\% C.L. on the  anomalous top Yukawa coupling is projected to be 5.6\% (1.7\%) at a 10 (30) TeV muon collider, which is comparable to the sensitivity of 2\% at the 100 TeV collider.
\end{abstract}

\section{Introduction}
Precision measurement and direct resonance searches are the two main ways to search for new physics at colliders.  In the first category, we usually perform the measurement at the pole masses of known particles such as the $Z$-boson and Higgs boson~\cite{ALEPH:2005ab,LHCHiggsCrossSectionWorkingGroup:2011wcg}. The precision that can be achieved  is limited by the systematic and statistical uncertainties. In the second method, we look for the peak in the invariant mass distribution or other kinematical variables of the decay products of new particles, which directly probes the energy scale of the mass of the new heavy particles. However, it has long been noticed that  any modifications to the  couplings predicted by the spontaneously broken gauge theories will lead to some kind of energy growing behaviors that violate tree-level unitarity~\cite{Cornwall:1974km} (see~\cite{Liu:2022alx} recently for the  on-shell derivation). Specific to $2\rightarrow 2$ scattering, this means that anomalous couplings  will lead to energy growing behaviors rather than a constant, as one would expect in the Standard Model (SM).  It is well-known that $W_LW_L\rightarrow W_LW_L$ scattering processes  will grow like $E^2$ in the high energy limit for non-standard  Higgs gauge boson couplings. On the other hand, $W_LW_L\rightarrow f \bar f$ processes will grow like $E^2$ for anomalous gauge-boson fermion couplings and anomalous triple gauge boson couplings (aTGCs). It will also grow linearly with $E$ in the case of non-Standard  Higgs Yukawa couplings. This has motivated the precision measurements at the hadron colliders~\cite{Farina:2016rws,Franceschini:2017xkh,Liu:2018pkg,Panico:2021vav} due to the high energy bins available at the LHC.

Recently, there is a growing  interest  in the high energy  muon collider~\cite{Delahaye:2019omf,MuonCollider:2022nsa} and active researches are currently being done to explore the physics potential~\cite{Chiesa:2020awd, Costantini:2020stv, Capdevilla:2020qel, Han:2020uid,Han:2020pif,Han:2020uak,Buttazzo:2020ibd,Yin:2020afe,Buttazzo:2020uzc,Huang:2021nkl,Liu:2021jyc,Capdevilla:2021rwo,Han:2021udl,Capdevilla:2021fmj,Han:2021kes,AlAli:2021let,Asadi:2021gah,Bottaro:2021snn,Qian:2021ihf,Chiesa:2021qpr,Liu:2021akf,Chen:2021pqi,Chen:2022msz,Cesarotti:2022ttv,deBlas:2022aow,Bao:2022onq,Homiller:2022iax,Forslund:2022xjq}. The attractiveness of a muon collider lies in its availability to reach high energy $\gtrsim 10$ TeV while keeping the systematical uncertainty under control. It can potentially achieve high integrated luminosity as follows:
\beq
L = \left(\frac{\sqrt{s_\mu}}{10\, \text{TeV}}\right)^2 \times 10\, \text{ab}^{-1}
\eeq
This will allow percent-level precision in the high energy bin of  $\gtrsim 10$ TeV for muon annihilation electroweak processes with final states of  di-fermions and di-bosons ~\cite{Chen:2022msz}, which is essentially probing the 100 TeV scale. Note that this is higher than the flavor physics scale in the composite Higgs scenarios~\cite{Csaki:2008zd,Keren-Zur:2012buf}.  The high energy muon collider can also be considered as a gauge boson collider due to the logarithmic growth of the electroweak gauge boson parton distribution functions~\cite{Costantini:2020stv,Han:2020uid,AlAli:2021let}. In this paper, we take $\mu^+\mu^- \rightarrow t \bar t \nu\bar\nu$ at high energy muon collider as an example to illustrate the extent to which the energy growing behaviors in the vector boson fusion subprocess  can help to measure  anomalous  couplings. We will employ the effective $W$-boson approximation (EWA)~\cite{Dawson:1984gx,Kunszt:1987tk,Borel:2012by} with the analytical formulae of the helicity amplitudes for the subprocess $W^+W^-\rightarrow t \bar{t}$  to analyze the energy scaling behavior of the signal significance. After the semi-analytical study, we will move on to study the prospects for the measurement of the top Yukawa coupling at the 10 TeV and 30 TeV muon collider, where we are focusing on the semi-leptonically decaying channel of the top quark pair.

%

The paper is organized as follows. In Section~\ref{sec:escal}, we perform a generic  energy scaling analysis of signal significance  for vector boson fusion processes in the presence of anomalous couplings at the high energy muon colliders. In particular, we give a detailed analysis for the $W^+ W^- \rightarrow t \bar t$ process and study the high energy and threshold behaviors for the helicity amplitudes in the presence of anomalous top Yukawa coupling, aTGCs and anomalous gauge-boson-fermions couplings. In Section~\ref{sec:yt}, we study the prospects on the  top Yukawa coupling measurement in the VBF production of $t \bar t$  at the high energy muon collider. For future possible study, we have also performed a brief analysis about the  top Yukawa coupling measurement in the VBF production of $t \bar t h$.  The results are presented in the Section~\ref{sec:result}. Section~\ref{sec:conclu} contains our conclusions.

\section{General Analysis of Weak Boson Fusion Processes }
\label{sec:escal}

In this section, we will study the energy scaling behavior of $S/\sqrt{B}$ and $S/B$ in the presence of anomalous couplings for the weak boson fusion  processes in the two particle final states at the high energy muon collider. We will focus on the hard scattering regime where the scattering angle is in the central region, i.e. $-\hat t\sim\hat s =\hat E^2$. We start from the analysis of the partonic processes $VV\rightarrow XY$ and then employ the Effective $W$-boson Approximation (EWA) \cite{Dawson:1984gx,Kunszt:1987tk,Borel:2012by} to analyze the energy scaling at the $\mu^+\mu^-$ collider.

\subsection{Energy Scaling Behavior in $W^+ W^- \rightarrow XX, ZZ \rightarrow XX, WZ \rightarrow XY$}
As a preliminary step to understanding the energy scaling behavior of processes at a muon collider, we consider the simpler problem of $VV \rightarrow XY$ where the $V$ stands for a $W $ or $Z$ boson and $X,Y$ can be any SM particles with electroweak charges such that the processes have non-zero tree-level contributions. We can later relate the results from this analysis to $\mu^+\mu^-$ cross sections by the Effective $W$ Approximation \cite{Borel:2012by,Kunszt:1987tk,Dawson:1984gx}. Restricting to 2 $\rightarrow$ 2 processes where the initial state contains two massive bosons, we can express our cross sections schematically in terms of amplitudes as:
\begin{equation}
\sigma_{int} \sim \frac{\mM_{\rm SM}\mM_{\delta_i}}{\hat{E}^2}; \quad \sigma_{\rm SM} \sim \frac{\mM_{\rm SM}^2}{\hat{E}^2}, \label{crosseq}
\end{equation}
where $\mM_{\rm SM}$ refers to the SM amplitude and $\mM_{\delta_i}$ refers to amplitudes containing BSM physics.  As only the energy scaling is concerned here, we also neglect the possible phase of the amplitudes. Note that we also study the hard scattering regime which is away from the possible scattering angle singularities (mainly from $t$-channel or $u$-channel).
Then given our processes, $\mM_{\delta_i}$ is linear in the anomalous couplings $\delta_i$ and we see that $\sigma_{int}$ is the interference term. We start from the analysis by assuming that we can exactly measure the helicities of the initial bosons and final state particles, so we are really considering:
\begin{equation}
\frac{S^{h_1...h_4}}{\sqrt{B^{h_1...h_4}}}. \label{sbhel}
\end{equation}
where the signal in the helicity configuration $S^{h_1...h_4}$ is linear in the coupling modifier $\delta_i$. In what follows, we only consider the SM process $VV\rightarrow XY$ as our dominant background. It is straightforward to see that under our simplified assumption,  for the case where statistical error dominates, the dependence on the SM amplitude of the statistical significance cancels out:
\begin{equation}
\frac{S^{h_1...h_4}}{\sqrt{B^{h_1...h_4}}} \sim \frac{\sigma^{h_1 \cdots h_4}_{int}}{\sqrt{\sigma^{h_1 \cdots h_4}_{\rm SM}}} \sim\frac{\mM^{h_1\cdots h_4}_{\delta_i}}{\hat{E}} \label{sbhelstat}
\end{equation}
Note that we have neglected all the constant factors, like the integrated luminosity.
Then we can see that in order for the significance to  grow with energy, $\mM_{\delta_i}$ must be at least quadratic in $\hat E$. This is certainly true for the Higgs gauge boson coupling modification in the vector boson scattering processes $ V_L V_L \rightarrow V_L V_L$ and for the anomalous gauge boson fermion coupling in the  $V_L V_L \rightarrow f \bar{f} $ processes. However for the top Yukawa coupling, we only have linear energy growing behavior and we expect that the significance stays constant as the bin energy increases. This does not mean the high energy bins are completely irrelevant, as one can still improve the significance by combing all the energy bins.


In reality, we cannot measure the helicities of the final states exactly and there is always contamination from other helicity categories. At the muon collider, it will likely be difficult to determine the initial gauge boson helicities, especially for the $W^\pm$  bosons. We now consider the inclusive case, where we sum over the cross sections from all the helicity configurations for the initial and final states.  In this fully inclusive case, the statistical significance scales like:
\begin{equation}
\frac{S}{\sqrt{B}} \sim \frac{\sum_{h_1...h_4}\sigma_{int}^{h_1...h_4}}{\sqrt{\sum_{h_1...h_4}\sigma_{SM}^{h_1...h_4}}} \sim \frac{1}{\hat{E}} \sum_{h_1\cdots h_4} \mM^{h_1\cdots h_4}_{\rm SM}\mM^{h_1\cdots h_4}_{\delta_i}\label{sbnohelstat}
\end{equation}
where we have used the fact the inclusive SM cross section has the following energy scaling:
\beq
\sum_{h_1...h_4}\sigma_{\rm SM}^{h_1...h_4} \sim \frac{1}{\hat E^2}
\eeq
We can see that in order for the significance to increase with energy, not only should the BSM helicity amplitude $\mM_{\delta_i}^{h_1\cdots h_4}$ grow as $\hat E^2$, but the corresponding linearly mixing term $\mM_{\rm SM}^{h_1 \cdots h_4}$ should also stay constant as the energy increases~\footnote{When taking into account the angular distributions of the decayed products of  final particles, the requirement may be relaxed as different helicity configurations of $XY$ can interfere with each other~\cite{Azatov:2017kzw,Panico:2017frx}.}.  Before studying the energy scaling of the weak boson parton luminosity in detail, we comment on the systematic uncertainty. If the systematic error dominates, the signal significance becomes:
\begin{equation}
\frac{S^{h_1...h_4}}{B^{h_1...h_4}} \sim \frac{\sigma^{h_1\cdots h_4}_{int}}{\sigma^{h_1 \cdots h_4}_{\rm SM}} \sim  \frac{\mM^{h_1\cdots h_4}_{\delta_i}}{\mM^{h_1 \cdots h_4}_{\rm SM}}. \label{sbhelsys}
\end{equation}
while for the inclusive case, it reads:
\begin{equation}
\frac{S}{B} = \frac{\sum_{h1...h4}\sigma_{int}^{h_1...h_4}}{\sum_{h_1\cdots h_4}\sigma_{\rm SM}^{h_1...h_4}} \sim  \sum_{h_1\cdots h_4} \mM^{h_1\cdots h_4}_{\rm SM}\mM^{h_1\cdots h_4}_{\delta_i} \label{sbnohelsys}
\end{equation}
In the exclusive case, since the SM helicity amplitudes $\mM_{\rm SM}^{h_1 \cdots h_4}$ are at most a constant for the $2\rightarrow 2$ processes, any energy growing behavior in the BSM amplitude $\mM_{\delta_i}^{h_1 \cdots h_4}$ will lead to enhancement of the signal significance at high energy bins. This is especially the case at the hadron colliders like the LHC, as one generally has large systematic errors ranging from a few percent to tens of percents. For the inclusive case, similar to the statistical uncertainty dominance, we need both $\mM_{\delta_i}^{h_1 \cdots h_4}$ to increase with energy and $\mM_{\rm SM}^{h_1 \cdots h_4}$ to not decrease too quickly.   



\subsection{Anatomy of $W^+ W^- \rightarrow t \bar t$}
\begin{figure}[t]
\centerline{\includegraphics[width=300pt]{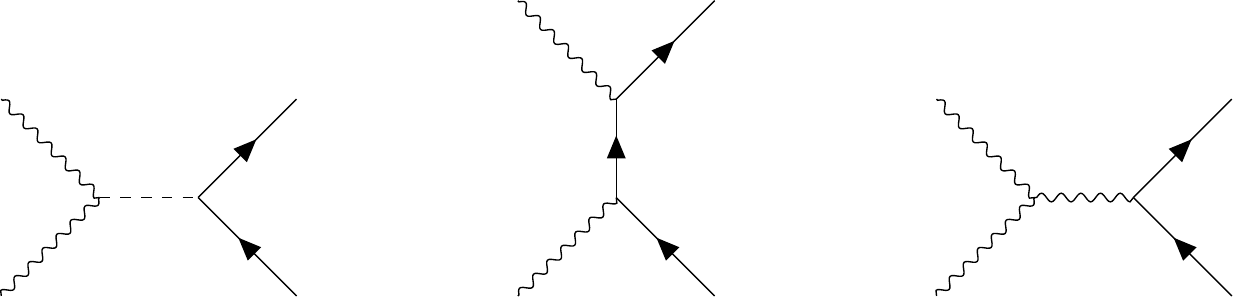}}
\caption{Tree-level diagrams for $W^+ W^- \rightarrow  t \bar t$. The gauge boson propagator in the last diagrams can be either $Z$ or $\gamma$. }\label{feyndiag}
\end{figure}
\begin{table}[t]
\caption{High energy limit of the  Helicity amplitude  for $W^+ W^- \rightarrow t \bar t$ with $h_t - h_{\bar t} = \mp 1$. Here $m_{\rm SM}$ denotes $m_W, m_t, m_h$. }
\begin{center}
\tabcolsep 4pt \begin{tabular}{|c|c|c|c|c|c|c|c|c|}
\hline
$(h_t \, h_{\bar t})$ & $(h_{W^+} h_{W^-})$& $\widetilde \mM_{h_{W^+} h_{W^-};h_t \, h_{\bar t}}^{\rm SM}$ & $ \widetilde\mM_{h_{W^+} h_{W^-};h_t \, h_{\bar t}}^{\rm BSM}$  \\
  \hline
\multirow{5}{*}{$(-\frac12\ \frac12)$}        & (+1 -1),(-1 +1)   & $i \frac{g^2}{-1+\cos\theta}$ & $\mO(\delta_{Wtb})$  \\
     &  (+1 +1),(-1 -1)  &  $\mO(\frac{m_{\rm SM}^2}{\hat E^2})$ & $i \frac{g^2 \hat E^2 (3 \lambda_Z + 4 s_W^2 (\lambda_\gamma-\lambda_Z))}{6\sqrt{2} m_{W}^2}$   \\
    & (+1 0),(0 -1)   & $\mathcal{O} (\frac{m_{\rm SM}}{\hat E})$& $\mO\left(\frac{\hat E}{m_{\rm SM}} (\delta g_1^Z, \delta \kappa_{Z,\gamma},\delta_{Wtb},\delta_{Zt_L},\lambda_{Z,\gamma})\right)$\\
     & (-1 0),(0 +1)   &$\mathcal{O} (\frac{m_{\rm SM}}{\hat E})$ &$\mO\left(\frac{\hat E}{m_{\rm SM}} (\delta g_1^Z, \delta \kappa_{Z,\gamma},\delta_{Wtb},\delta_{Zt_L},\lambda_{Z,\gamma})\right)$  \\
      & (0 0)&$  i \frac{3 g^2 + g^{\prime 2}}{6\sqrt{2}}$  &$  i \frac{g^2 \hat E^2}{6\sqrt{2} m_{W}^2}\left((-3+4s_W^2) (\delta \kappa_Z + \delta_{Zt_L}) + 6 \delta_{Wtb}- 4 s_W^2  \delta \kappa_\gamma)\right)$   \\
\hline
\end{tabular}
\vspace{0.5cm}
\begin{tabular}{|c|c|c|c|c|c|c|c|c|}
\hline
$(h_t \, h_{\bar t})$ & $(h_{W^+} h_{W^-})$& $\widetilde \mM_{h_{W^+} h_{W^-};h_t \, h_{\bar t}}^{\rm SM}$ & $ \widetilde\mM_{h_{W^+} h_{W^-};h_t \, h_{\bar t}}^{\rm BSM}$  \\
  \hline
\multirow{5}{*}{$(\frac12\, -\frac12)$}        & (+1 -1),(-1 +1)   &  $\mO(\frac{m_{\rm SM}^2}{\hat E^2})$ &   $\mO(\frac{m_{\rm SM}^2}{\hat E^2})$ \\
     &  (+1 +1),(-1 -1)  &  $\mO(\frac{m_{\rm SM}^2}{\hat E^2})$ & $i \frac{\sqrt{2 }g^2 s_W^2  \hat E^2     (\lambda_\gamma-\lambda_Z)}{3 m_{W}^2}$  \\
    & (+1 0),(0 -1)   &  $\mO(\frac{m_{\rm SM}}{\hat E})$  & $\mO\left(\frac{\hat E}{m_{\rm SM}} (\delta g_1^Z, \delta\kappa_{Z,\gamma},\delta_{Zt_R},\lambda_{Z,\gamma})\right)$ \\
     & (-1 0),(0 +1)   & $\mO(\frac{m_{\rm SM}}{\hat E})$  &  $\mO\left(\frac{\hat E}{m_{\rm SM}} (\delta g_1^Z, \delta\kappa_{Z,\gamma},\delta_{Zt_R},\lambda_{Z,\gamma})\right)$  \\
      & (0 0)& $ i \frac{ \sqrt{2} g^{\prime 2}}{3} + i \frac{2\sqrt{2} m_t^2}{v^2(-1+\cos\theta)}$   & $  i \frac{\sqrt{2}g^2   s_W^2 \hat E^2}{3 m_{W}^2}\left( \delta \kappa_Z - \delta \kappa_\gamma + \delta_{Zt_R}\right)$ \\
\hline
\end{tabular}
\vspace{0.5cm}
\end{center}
\label{tab:hche1}
\end{table}

\begin{table}[t]
\caption{High energy limit of the  Helicity amplitude  for $W^+ W^- \rightarrow t \bar t$ with $h_t - h_{\bar t} = 0$. Here $m_{\rm SM}$ denotes $m_W, m_t, m_h$. }
\begin{center}
\begin{tabular}{|c|c|c|c|c|c|c|c|c|}
\hline
$(h_t \, h_{\bar t})$ & $(h_{W^+} h_{W^-})$& $\widetilde \mM_{h_{W^+} h_{W^-};h_t \, h_{\bar t}}^{\rm SM}$ & $ \widetilde\mM_{h_{W^+} h_{W^-};h_t \, h_{\bar t}}^{\rm BSM}$  \\
  \hline
\multirow{5}{*}{$(-\frac12\ -\frac12)$}        & (+1 -1),(-1 +1)   & $\mathcal{O} (\frac{m_{\rm SM}}{\hat E} )$ &$\mathcal{O} (\frac{m_{\rm SM}}{\hat E} \delta_{Wtb})$ \\
     &  (+1 +1)  & $\mO(\frac{m_{\rm SM}^3}{\hat E^3})$  &  $\mO\left(\frac{\hat E}{m_{\rm SM}} \lambda_{Z,\gamma}\right)$   \\
          & (-1 -1)  &  $\mO(\frac{m_{\rm SM}}{\hat E})$  & $\mO\left(\frac{\hat E}{m_{\rm SM}} \lambda_{Z,\gamma}\right)$   \\
    & (+1 0),(0 -1)   & $\mathcal{O} (\frac{m_{\rm SM}^2}{\hat E^2})$& $\mO\left(\delta g_1^Z, \delta \kappa_{Z,\gamma},\delta_{Wtb},\lambda_{Z,\gamma},\delta_{Zt_L},\delta_{Zt_R}\right)$\\
     & (-1 0)   &$i g^2 \frac{m_t}{m_{W}(-1+\cos\theta)}$ &$\mO\left(\delta g_1^Z,\delta \kappa_{Z,\gamma},\delta_{Wtb},\lambda_{Z,\gamma},\delta_{Zt_L},\delta_{Zt_R}\right)$  \\
          & (0 +1)   &$\mathcal{O} (\frac{m_{\rm SM}^2}{\hat E^2})$ &$\mO\left(\delta g_1^Z, \delta \kappa_{Z,\gamma},\delta_{Wtb},\lambda_{Z,\gamma},\delta_{Zt_L},\delta_{Zt_R}\right)$  \\
      & (0 0)& $\mO(\frac{m_{\rm SM}}{\hat E})$  &$ - i \frac{g^2 m_t\hat E }{4 m_{W}^2}\left( \delta_{hWW} + \delta_{tth} +\mO(\delta \kappa_{Z,\gamma},\delta_{Zt_L}, \delta_{Zt_R})\right)$   \\
\hline
\end{tabular}
\vspace{0.5cm}
\begin{tabular}{|c|c|c|c|c|c|c|c|c|}
\hline
$(h_t \, h_{\bar t})$ & $(h_{W^+} h_{W^-})$& $\widetilde \mM_{h_{W^+} h_{W^-};h_t \, h_{\bar t}}^{\rm SM}$ & $ \widetilde\mM_{h_{W^+} h_{W^-};h_t \, h_{\bar t}}^{\rm BSM}$  \\
  \hline
\multirow{5}{*}{$(\frac12\ \frac12)$}        & (+1 -1),(-1 +1)   & $\mathcal{O} (\frac{m_{\rm SM}}{\hat E})$ &$\mathcal{O} (\frac{m_{\rm SM}}{\hat E} \delta_{Wtb})$ \\
     &  (+1 +1)  &$\mO(\frac{m_{\rm SM}}{\hat E})$   & $\mO(\frac{\hat E}{m_{\rm SM}} \lambda_{Z,\gamma})$   \\
          & (-1 -1)  & $\mO(\frac{m_{\rm SM}^3}{\hat E^3})$   & $\mO(\frac{\hat E}{m_{\rm SM}} \lambda_{Z,\gamma})$   \\
    & (+1 0),(0 -1)   & $\mathcal{O} (\frac{m_{\rm SM}^2}{\hat E^2})$&$\mO\left(\delta g_1^Z, \delta \kappa_{Z,\gamma},\delta_{Wtb},\lambda_{Z,\gamma},\delta_{Zt_L},\delta_{Zt_R}\right)$\\
     & (-1 0)   & $\mathcal{O} (\frac{m_{\rm SM}^2}{\hat E^2})$ &$\mO\left(\delta g_1^Z, \delta \kappa_{Z,\gamma},\delta_{Wtb},\lambda_{Z,\gamma},\delta_{Zt_L},\delta_{Zt_R}\right)$ \\
          & (0 +1)   & $i g^2 \frac{m_t}{m_{W}(1-\cos\theta)}$&$\mO\left(\delta g_1^Z,\delta \kappa_{Z,\gamma},\delta_{Wtb},\lambda_{Z,\gamma},\delta_{Zt_L},\delta_{Zt_R}\right)$  \\
      & (0 0)& $\mO(\frac{m_{\rm SM}}{\hat E})$  &$  i \frac{g^2  m_t \hat E}{4 m_W^2}\left( \delta_{hWW} + \delta_{tth} +\mO(\delta \kappa_{Z,\gamma},\delta_{Zt_L}, \delta_{Zt_R})\right)$   \\
\hline
\end{tabular}
\vspace{0.5cm}
\end{center}
\label{tab:hche2}
\end{table}

In this subsection, we focus on the VBF production of the top pair and study in detail the helicity amplitudes of the subprocess $W^+ W^- \rightarrow t \bar t$ in the presence of anomalous couplings. The relevant Feynman diagrams are shown in Fig. \ref{feyndiag}. For completeness and also for future possible studies, we also include the anomalous triple gauge boson couplings (aTGCs), the gauge boson fermion couplings and Higgs gauge boson coupling.  The full formulae and the conventions are presented in the Appendix~\ref{app:hcamp}. Here we discuss their high energy and threshold behaviors. We start from the high energy hard scattering limit and consider the central region, where $1 \pm \cos\theta$ is large enough to justify our expansion. As before, we denote $\hat E = \sqrt{\hat s}$. The results for the helicity-conserving configurations of the top quarks, i.e. ($h_t, h_{\bar t} = (\pm \frac 12, \mp \frac 12)$) are listed in Table~\ref{tab:hche1}, while the results for the helicity-violating  configurations i.e. ($h_t, h_{\bar t} = (\pm \frac 12, \pm \frac 12)$) are presented in Table~\ref{tab:hche2}. The energy scaling for the helicity partonic cross section and the exclusive statistical significance is given in Table~\ref{tab:xshe}. Several comments are in order. First, for the SM helicity amplitudes, only the following helicity configurations survive in the high energy limit:
\beq
 (h_{W^+}, h_{W^-}, h_t,h_{\bar t}) = (\pm 1, \mp 1,- \frac 12,  \frac 12), \quad (0, 0, \mp \frac 12,  \pm \frac 12), \quad (- 1, 0, - \frac 12,  - \frac 12), \qquad (0, 1, \frac 12,   \frac 12)
\eeq
The results can be understood by using the Goldstone equivalence theorem and by working in the electroweak-symmetry-unbroken phase of the SM where the Goldstone scalars $\phi_{\pm}$ appear as external states and the SM gauge bosons and top quarks are massless particles. For the longitudinal $W^\pm$ bosons processes, we can see that the $SU(2)_L\times U(1)_Y$ quantum numbers of the top quarks appear in the helicity amplitudes:
\beq
T_3^L (t_{L,R}) g^2 + Y (t_{L,R}) g^{\prime 2}
\eeq
where $T_3^L$ is the third weak isospin generator and $Y$ is the hypercharge. The presence of the SM top Yukawa coupling squared term $m_t^2/v^2$ associated with $t$-channel pole in the $(0,0,\frac 12, -\frac 12)$ configuration is due to  left-handed bottom quark exchange diagram in the $\phi^+\phi^- \rightarrow t \bar t$ process. Note  that if the bottom quark mass were not set to zero in our calculation, there would be a similar term with $m_b^2/v^2$  in the $(0,0,-\frac 12, \frac 12)$ configuration. Following this reasoning, we can understand the processes involving only one longitudinal gauge bosons $W^\pm \phi^\mp \rightarrow t \bar t$.

\begin{table}[t]
\small
\caption{Energy scaling  for  cross sections and statistical signal significance of  $W^+W^- \rightarrow t \bar{t} $ in different helicity categories with different anomalous couplings.  The results for $\delta_{hWW}$ has the same behavior as $\delta_{tth}$ and therefore are not shown here.}
\label{tab:xshe}
\hskip-1.75cm 
\begin{tabular}{|c||P{0.5cm}|P{0.75cm}|P{0.55cm}|P{0.7cm}|P{0.7cm}|P{0.5cm}|P{0.5cm}|P{0.5cm}||P{0.65cm}|P{0.65cm}|P{0.65cm}|P{0.65cm}|P{0.65cm}|P{0.65cm}|P{0.65cm}|P{0.65cm}|P{0.65cm}|P{0.65cm}|P{0.65cm}|P{0.65cm}|}
\hline
($h_{W^+}$, $h_{W^-}$, $h_t$, $h_{\bar{t}}$) & $\hat{\sigma}_{\rm SM}$ & $\hat{\sigma}_{\delta_{tth}}$ & $\hat{\sigma}_{\lambda_{Z, \gamma}}$ & $\hat{\sigma}_{\delta \kappa_{Z,\gamma}}$& $\hat{\sigma}_{\delta_{Wtb}}$&$\hat{\sigma}_{Z t_L}$&$\hat{\sigma}_{Zt_R}$ & $\hat{\sigma}_{\delta g_1^Z}$ &$\frac{S_{\delta_{tth}}}{\sqrt{B}}$ & $\frac{S_{\lambda_{Z,\gamma}}}{\sqrt{B}}$ & $\frac{S_{\delta \kappa_{Z,\gamma}}}{\sqrt{B}}$ & $\frac{S_{\delta_{Wtb}}}{\sqrt{B}}$ &  $\frac{S_{\delta_{Z t_L}}}{\sqrt{B}}$ & $\frac{S_{\delta_{Z t_R}}}{\sqrt{B}}$ & $\frac{S_{\delta g_1^Z}}{\sqrt{B}}$\\
\hline
(0, 0, $-$, $+$) & $\frac{1}{\hat{E}^2}$ & $\times$ & $\times$& $\hat{E}^0$ & $\hat{E}^0$& $\hat{E}^0$ & $\times$ & $\times$ & $\times$ &$\times$& $\pmb{\hat E}$ &$\pmb{\hat{E}}$ & $\pmb{\hat{E}}$ & $\times$ & $\times$ \\
(0, 0, $+$, $-$) & $\frac{1}{\hat{E}^2}$ & $\times$ & $\times$ &$\hat E^0$& $\times$ & $\times$ & $\hat{E}^0$&$\times$&$\times$ & $\times$ & {$ \pmb{\hat E}$} & $\times$ & $\times$ & $\pmb{\hat{E}}$ & $\times$ \\
(0, 0, $\mp$, $\mp$) & $\frac{1}{\hat{E}^4}$ & $\frac{1}{\hat{E}^2}$ &$\times$&$\frac{1}{\hat{E}^2}$&$\times$ &$\frac{1}{\hat{E}^2}$ & $\frac{1}{\hat{E}^2}$&$\times$ &$\hat{E}^0$ &$\times$& $\hat{E}^0$ & $\times$ & $\hat{E}^0$ & $\hat{E}^0$ & $\times$ \\ [5pt]
\hline
$\begin{array}{cc}(0,  +, +, +)\\(-,0,-,-) \end{array}$ &  $\frac{1}{\hat{E}^2}$ & $\times$ &$\frac{1}{\hat{E}^2}$ &$\frac{1}{\hat{E}^2}$&$\frac{1}{\hat{E}^2}$&$\frac{1}{\hat{E}^2}$ &$\frac{1}{\hat{E}^2}$ &$\frac{1}{\hat{E}^2}$ & $\times$&$\frac{1}{\hat E}$ & $\frac{1}{\hat{E}}$ & $\frac{1}{\hat{E}}$ &  $\frac{1}{\hat{E}}$ & $\frac{1}{\hat{E}}$  & $\frac{1}{\hat{E}}$ \\ [5pt]
$\begin{array}{cc}(0, +, -, -)\\ (-, 0, +, + )\end{array}$&  $\frac{1}{\hat{E}^6}$ & $\times$ & $\frac{1}{\hat{E}^4}$ & $\frac{1}{\hat{E}^4}$ & $\frac{1}{\hat{E}^4}$ & $\frac{1}{\hat{E}^4}$ & $\frac{1}{\hat{E}^4}$ & $\frac{1}{\hat{E}^4}$&$\times$ & $\frac{1}{\hat{E}}$  & $\frac{1}{\hat{E}}$  & $\frac{1}{\hat{E}}$  & $\frac{1}{\hat{E}}$  & $\frac{1}{\hat{E}}$  & $\frac{1}{\hat{E}}$ \\ [5pt]
$\begin{array}{cc}(+, 0, -, -)\\ (0, -, -, -) \end{array}$ & $\frac{1}{\hat{E}^6}$ & $\times$ & $\frac{1}{\hat{E}^4}$ & $\frac{1}{\hat{E}^4}$& $\frac{1}{\hat{E}^4}$&$\frac{1}{\hat{E}^4}$&$\frac{1}{\hat{E}^4}$ & $\frac{1}{\hat{E}^4}$ &  $\times$ & $\frac{1}{\hat{E}}$  & $\frac{1}{\hat{E}}$  & $\frac{1}{\hat{E}}$  & $\frac{1}{\hat{E}}$  & $\frac{1}{\hat{E}}$  & $\frac{1}{\hat{E}}$\\[5pt]
$\begin{array}{cc}(0, -, +, +)\\ (+, 0, +, +)\end{array}$& $\frac{1}{\hat{E}^6}$ & $\times$ & $\frac{1}{\hat{E}^4}$ & $\frac{1}{\hat{E}^4}$& $\frac{1}{\hat{E}^4}$&$\frac{1}{\hat{E}^4}$&$\frac{1}{\hat{E}^4}$ & $\frac{1}{\hat{E}^4}$ & $\times$ & $\frac{1}{\hat{E}}$  & $\frac{1}{\hat{E}}$  & $\frac{1}{\hat{E}}$  & $\frac{1}{\hat{E}}$  & $\frac{1}{\hat{E}}$  & $\frac{1}{\hat{E}}$ \\[5pt]
$\begin{array}{cc}(+, 0, -, +)\\ (0, -, -, +)\end{array}$ & $\frac{1}{\hat{E}^4}$ & $\times$ & $\frac{1}{\hat{E}^2}$ & $\frac{1}{\hat{E}^2}$ &$\frac{1}{\hat{E}^2}$ & $\frac{1}{\hat{E}^2}$ & $\times$ & $\frac{1}{\hat{E}^2}$ &$\times$& $\hat{E}^0$ & $\hat{E}^0$ &  $\hat{E}^0$ & $\hat{E}^0$   & $\times$ & $\hat{E}^0$   \\[5pt]
$\begin{array}{cc}(-, 0, -, +)\\ (0, +, -, +) \end{array} $& $\frac{1}{\hat{E}^4}$ & $\times$ &$\frac{1}{\hat{E}^2}$ &$\frac{1}{\hat{E}^2}$&$\frac{1}{\hat{E}^2}$&$\frac{1}{\hat{E}^2}$ & $\times$ & $\frac{1}{\hat{E}^2}$ & $\times$ & $\hat{E}^0$ & $\hat{E}^0$ &  $\hat{E}^0$ & $\hat{E}^0$   & $\times$ & $\hat{E}^0$  \\ [5pt]
 $\begin{array}{cc}(+, 0, +, -) \\ (0, -, +, -)\end{array}$& $\frac{1}{\hat{E}^4}$ & $\times$ &$\frac{1}{\hat{E}^2}$  &$\frac{1}{\hat{E}^2}$ &$\times$&$\times$& $\frac{1}{\hat{E}^2}$ & $\frac{1}{\hat{E}^2}$ & $\times$& $\hat{E}^0$ & $\hat{E}^0$ & $\times$ & $\times$ & $\hat{E}^0$   & $\hat{E}^0$  \\ [5pt]
$\begin{array}{cc}(-, 0, +, -)\\ (0, +, +, -)\end{array}$ & $\frac{1}{\hat{E}^4}$ & $\times$& $\frac{1}{\hat{E}^2}$ &$\frac{1}{\hat{E}^2}$ &$\times$&$\times$ & $\frac{1}{\hat{E}^2}$ & $\frac{1}{\hat{E}^2}$ & $\times$ & $\hat{E}^0$ & $\hat{E}^0$ & $\times$ & $\times$ & $\hat{E}^0$   & $\hat{E}^0$  \\
\hline
($\pm$, $\mp$, $-$, +) & $\frac{1}{\hat{E}^2}$ &$\times$  & $\times$&$\times$& $\frac{1}{\hat{E}^2}$& $\times$ &$\times$ & $\times$ &$\times$ & $\times$ & $\times$ & $\frac{1}{\hat{E}}$& $\times$ & $\times$ & $\times$\\
($+$, $+$, $-$, $-$) & $\frac{1}{\hat{E}^8}$ & $\times$ &$\frac{1}{\hat{E}^4}$ & $\times$ & $\times$& $\times$&$\times$ & $\times$& $\times$ & $\hat{E}^0$  & $\times$ & $\times$ & $\times$ & $\times$ & $\times$ \\ 
($-$, $-$, $-$, $-$) & $\frac{1}{\hat{E}^4}$ &$\times$ &$\frac{1}{\hat{E}^2}$ &$\times$ & $\times$& $\times$& $\times$& $\times$& $\times$ & $\hat{E}^0$  & $\times$ & $\times$ & $\times$ & $\times$ & $\times$ \\ 
($\pm$, $\mp$, $-$, $-$) & $\frac{1}{\hat{E}^4}$ & $\times$ & $\times$&$\times$&$\frac{1}{\hat{E}^4}$&$\times$& $\times$&$\times$&$\times$&$\times$&$\times$& $\frac{1}{\hat{E}^2}$   & $\times$ & $\times$ & $\times$ \\
($\pm$, $\mp$, $+$, $+$) & $\frac{1}{\hat{E}^4}$ & $\times$ & $\times$&$\times$&$\frac{1}{\hat{E}^4}$&$\times$& $\times$& $\times$& $\times$&$\times$&$\times$&$\frac{1}{\hat{E}^2}$ & $\times$ & $\times$ & $\times$\\
($+$, $+$, +, $+$) & $\frac{1}{\hat{E}^4}$ & $\times$ & $\frac{1}{\hat{E}^2}$& $\times$ &$\times$ & $\times$& $\times$&$\times$ & $\times$ & $\hat{E}^0$  & $\times$ & $\times$ & $\times$ & $\times$ & $\times$ \\ 
($-$, $-$, +, $+$) & $\frac{1}{\hat{E}^8}$ &$\times$ & $\frac{1}{\hat{E}^4}$ &$\times$ & $\times$& $\times$&$\times$ & $\times$ & $\times$ & $\hat{E}^0$  & $\times$ & $\times$ & $\times$ & $\times$ & $\times$ \\ 
($\pm$, $\pm$, $-$, +) & $\frac{1}{\hat{E}^6}$ & $\times$ &$\frac{1}{\hat E^2}$ &$\times$&$\times$& $\times$ & $\times$ & $\times$ &$\times$&$\pmb{\hat E}$&$\times$&$\times$ & $\times$ & $\times$ & $\times$ \\
($\pm$, $\pm$, +, $ -$) &  $\frac{1}{\hat{E}^6}$  & $\times$ &$\frac{1}{\hat E^2}$ &$\times$&$\times$& $\times$ &$\times$&$\times$&$\times$&$\pmb{\hat E}$&$\times$&$\times$ & $\times$ & $\times$ & $\times$ \\[7pt]
\hline
\end{tabular}
\end{table}

Secondly, we can see from the Table~\ref{tab:hche1} that for the anomalous triple gauge boson couplings $\delta \kappa_{Z,\gamma}$ in the  ($\mp\frac12, \pm \frac12$) top quark pair helicities and the anomalous top quark electroweak coupling $\delta_{Wtb}, \delta_{Zt_L} (\delta_{Zt_R})$  in the  ($-\frac12,  \frac12$) ( ($\frac12,  -\frac12$)) top quark pair helicities, the helicity amplitudes from longitudinal gauge bosons  scale like $\hat E^2$, while the SM contributions stay constant in the high energy limit. As discussed above, this means that for both the exclusive channel with all the helicities of the particles fully measured and the inclusive channel where all the helicity configurations are included, the statistical significance scales like $\hat E$, which results in larger sensitivity for higher energy bins. However, for the modification of the top Yukawa coupling $\delta_{tth}$ in the high energy limit~\footnote{Likewise for the anomalous Higgs gauge boson coupling $\delta_{hWW}$, as only the combination of $\delta_{tth} + \delta_{hWW} $ appears in the helicity amplitudes at linear order.}, the helicity amplitude only grows linearly as $\hat E$ in the $(0,0,\mp \frac12, \mp \frac12)$ helicity configuration and the SM contribution decreases like $1/\hat E$. This in turn leads to the constant behavior for the statistical significance in the exclusive channel and decreasing statistical significance as  $\mO(1/\hat E)$ in the inclusive channel. This means that in the realistic case at the muon collider, the sensitivity on the top Yukawa coupling from the electroweak top pair production would mostly come from low energy bins. The high energy muon collider benefits us from the growth of the  VBF cross sections, i.e. the enhancement of the vector boson parton luminosity.  We finally note that for the case of systematical uncertainty dominance, the significance grows as energy increases for all anomalous couplings in the exclusive channel.  For the fully inclusive channel, the significance grows as $\hat E^2$ for the anomalous couplings $\delta \kappa_{Z,\gamma}, \delta_{Wtb}, \delta_{Z t_L}, \delta_{Zt_R}$, but stays constant for aTGC $\lambda_{Z,\gamma}$ and the anomalous top Yukawa coupling $\delta_{tth}$.


Now we examine the threshold behavior of top quark electroweak pair production. We expand the helicity amplitudes in terms of the top quark velocity $\beta_t$ around the $\sqrt{s} \sim 2 m_t$. For simplicity, we also keep only the leading power of $m_{W,Z}^2/m_t^2$. The results are presented in Table~\ref{tab:hcth1} for the helicity configurations $(h_{t}, h_{\bar t}) = (\mp \frac12, \pm \frac 12)$ and listed in Table~\ref{tab:hcth2} for the helicity configurations $(h_{t}, h_{\bar t}) = (\mp \frac12, \mp \frac 12)$. We can see from the tables that  all the SM helicity amplitudes arise at the zeroth order of  top quark velocity $\beta_t^0$ except the helicity configurations for $(h_{W^+}, h_{W^-}) = (\pm 1, \mp 1)$ 
as they arise from the $J \geq 2$ partial waves.  We also find that for the processes involving the longitudinal $W$ bosons, there is an additional factor of $m_t/m_W$ enhancement for each longitudinal mode. For the anomalous TGCs $\delta \kappa_{Z,\gamma}$ in the  helicity configuration of the longitudinal $W^\pm$ bosons and $\lambda_{Z,\gamma}$ in the helicity configurations $(h_{W^+}, h_{W^-}) = (\pm 1, \pm 1))$, the amplitudes at threshold are enhanced by $m_t^2/m_W^2$ for all the helicity configurations of top quark pair. Since the SM contribution to amplitudes of  $(h_{W^+}, h_{W^-} )= (\pm 1, \pm 1))$ at threshold are not suppressed, it provides an interesting possibility to measure  aTGCs $\lambda_{Z,\gamma}$, which we leave for future studies. For the top Yukawa coupling modification $\delta_{tth}$, its leading contribution to the longitudinal $W^\pm$ gauge boson arises at order $\beta_t$, which means that the linear BSM helicity cross sections arise at $\beta_t^2$~\footnote{The extra $\beta_t$ comes in because  the final two-body phase space has linear dependence on the velocity of the top quark.}. The statistical significance will scale like $\beta_t^{3/2}$ in the small $\beta_t$ approximation and we need to have sizable top quark velocity to achieve maximal sensitivity. 

\begin{table}[t]
\caption{Threshold behaviors  of the  Helicity amplitude  for $W^+ W^- \rightarrow t \bar t $ with $h_t - h_{\bar t}= \mp 1$. Here we keep the leading terms in the top velocity $\beta_t$ expansion and  $\frac{m_{W,Z}^2}{m_t^2}$ expansion.}
\begin{center}
\begin{tabular}{|c|c|c|c|c|c|c|c|c|}
\hline
$(h_t \, h_{\bar t})$ & $(h_{W^+}\, h_{W^-})$& $\widetilde \mM_{h_{W^+} h_{W^-};h_t h_{\bar t}}^{\rm SM}$ & $ \widetilde\mM_{h_{W^+} h_{W^-};h_t h_{\bar t}}^{\rm BSM}$  \\
  \hline
\multirow{5}{*}{$(\mp\frac12\  \pm\frac12)$}        & (+1 -1),(-1 +1)   & $\mO(\beta_t)$ & $\mO(\beta_t \delta_{Wtb})$  \\
     &  (+1 +1),(-1 -1)  & $-i\frac{g^2}{2\sqrt{2}}$   & $i \frac{ g^2(3 \lambda_Z + 8 s_W^2(\lambda_\gamma-\lambda_Z))m_t^2}{3 \sqrt{2}m_W^2}$  \\
    & (+1 0),(0 -1)   & $-i \frac{g^2 m_t}{\sqrt{2} m_W}$& $\mO\left(\frac{m_t}{m_W} (\delta_{Wtb},\lambda_{Z,\gamma}, \delta g_1^Z, \delta \kappa_{Z,\gamma}, \delta_{Zt_L}, \delta_{Zt_R})\right)$ \\
        & (-1 0),(0 +1)   & $-i \frac{g^2 m_t}{\sqrt{2} m_W}$& $\mO\left(\frac{m_t}{m_W} (\lambda_{Z,\gamma},\delta g_1^Z, \delta \kappa_{Z,\gamma}, \delta_{Zt_L}, \delta_{Zt_R})\right)$ \\
	      & (0 0)&$ -i\frac{g^2m_t^2}{\sqrt{2}m_W^2}$  & $i \frac{ g^2(-3 ( \delta \kappa_Z+\delta_{Zt_L}) + 4 s_W^2(2\delta\kappa_Z- 2\delta\kappa_\gamma + \delta_{Zt_L} + \delta_{Z t_R}))m_t^2}{3 \sqrt{2}m_W^2}$  \\
\hline
\end{tabular}
\end{center}
\label{tab:hcth1}
\end{table}

\begin{table}[ht]
\caption{Threshold behaviors  of the  Helicity amplitude  for $W^+ W^- \rightarrow t \bar t $ with $h_t - h_{\bar t}= 0$. Here we keep the leading terms in the top velocity $\beta_t$ expansion and  $\frac{m_{W,Z}^2}{m_t^2}$ expansion.}
\begin{center}
\begin{tabular}{|c|c|c|c|c|c|c|c|c|}
\hline
$(h_t\, h_{\bar t})$ & $(h_{W^+} h_{W^-})$& $\widetilde \mM_{h_{W^+} h_{W^-};h_t h_{\bar t}}^{\rm SM}$ & $ \widetilde\mM_{h_{W^+} h_{W^-};h_t h_{\bar t}}^{\rm BSM}$  \\
  \hline
\multirow{5}{*}{$(-\frac12\ -\frac12)$}        & (+1 -1),(-1 +1)   & $\mathcal{O} (\beta_t)$ &$\mathcal{O} (\beta_t \delta_{Wtb})$ \\
     &  (+1 +1),(-1,-1)  & $i g^2\frac{\pm 2-\cos\theta}{4}$  & $i \frac{ g^2(3 \lambda_Z + 8 s_W^2(\lambda_\gamma-\lambda_Z))m_t^2 \cos\theta}{6 m_W^2} $   \\
    & (+1 0),(0 -1)   & $-i g^2\frac{m_t}{2 m_W}$& $\mO\left(\frac{m_t}{m_W} (\delta_{Wtb},\lambda_{Z,\gamma},\delta g_1^Z, \delta \kappa_{Z,\gamma},\delta_{Z t_L}, \delta_{Z t_R})\right)$\\
     & (-1 0),(0,+1)  &$-i g^2\frac{m_t}{2m_W}$ & $\mO\left(\frac{m_t}{m_W} (\lambda_{Z,\gamma},\delta g_1^Z, \delta \kappa_{Z,\gamma},\delta_{Z t_L}, \delta_{Z t_R})\right)$  \\ [5pt]
      & (0 0)& $-i \frac{g^2 m_t^2}{2m_W^2} \cos\theta$  &$\begin{array}{cc}i \frac{ g^2(-3 \delta  (\kappa_Z + \delta_{Zt_L}) + 4 s_W^2(2\delta\kappa_Z-2\delta\kappa_\gamma + \delta_{Zt_L} + \delta_{Zt_R}))m_t^2 \cos\theta}{6m_W^2}\\
      + i\frac{2g^2 m_t^4 \beta_t  (\delta_{tth} + \delta_{hWW})}{(m_h^2-4 m_t^2)m_W^2}\end{array}$   \\
\hline
\end{tabular}
\vspace{0.5cm}
\begin{tabular}{|c|c|c|c|c|c|c|c|c|}
\hline
$(h_t\, h_{\bar t})$ & $(h_{W^+} h_{W^-})$& $\widetilde \mM_{h_{W^+} h_{W^-};h_t h_{\bar t}}^{\rm SM}$ & $ \widetilde\mM_{h_{W^+} h_{W^-};h_t h_{\bar t}}^{\rm BSM}$  \\
  \hline
\multirow{5}{*}{$(\frac12\  \frac12)$}        & (+1 -1),(-1 +1)   & $\mathcal{O} (\beta_t)$ &$\mathcal{O} (\beta_t \delta_{Wtb})$ \\
     &  (+1 +1),(-1,-1)  & $i g^2\ \frac{\pm 2+\cos\theta}{4}$  & $-i \frac{ g^2(3 \lambda_Z + 8 s_W^2(\lambda_\gamma-\lambda_Z))m_t^2 \cos\theta}{6 m_W^2} $   \\
    & (+1 0),(0 -1)   & $i g^2\ \frac{m_t}{2m_W}$&$\mO\left(\frac{m_t}{m_W} (\delta_{Wtb},\lambda_{Z,\gamma},\delta g_1^Z, \delta \kappa_{Z,\gamma},\delta_{Z t_L}, \delta_{Z t_R})\right)$\\
     & (-1 0),(0,+1)  &$ig^2 \frac{ m_t}{2 m_W} $ &$\mO\left(\frac{m_t}{m_W} (\lambda_{Z,\gamma},\delta g_1^Z,\delta \kappa_{Z,\gamma},\delta_{Z t_L}, \delta_{Z t_R})\right)$  \\
      & (0 0)& $i \frac{g^2 m_t^2}{2m_W^2} \cos\theta$  &$\begin{array}{cc}-i \frac{ g^2(-3 \delta  (\kappa_Z + \delta_{Zt_L}) + 4 s_W^2(2\delta\kappa_Z-2\delta\kappa_\gamma + \delta_{Zt_L} + \delta_{Zt_R}))m_t^2 \cos\theta}{6m_W^2}\\
      - i\frac{2g^2 m_t^4 \beta_t  (\delta_{tth} + \delta_{hWW})}{(m_h^2-4 m_t^2)m_W^2}\end{array}$ \\
\hline
\end{tabular}
\vspace{0.5cm}
\end{center}
\label{tab:hcth2}
\end{table}

We finally comment on the scattering angle $\theta$ distribution, where $\theta$ is the polar angle between the outgoing top quark and incoming $W^+$ boson. As is well-known, there is a $t-$channel singularity in the cross section of this process, which can seen from the high energy limit in Table~\ref{tab:hche1} and appears in the helicity configuration $(h_{W^+}, h_{W^-}, h_t, h_{\bar t}) = (-1,+1,-\frac12, \frac 12)$. Note that to obtain the $\theta$ distribution for the helicity amplitudes, one needs to bring back the Wigner $d$-functions. For the $t$-channel singularity, the relevant functions are as follows:
\beq
\begin{split}
d_{-2,-1}^2 &= \frac12 \sin\theta (1 + \cos\theta),\qquad d_{2,-1}^2=- \frac12 \sin\theta (1 - \cos\theta)\\
\end{split}
\eeq
We can see that for other helicity configuration $(h_{W^+}, h_{W^-}, h_t, h_{\bar t}) = (+1,-1,-\frac12, \frac 12)$, the $t$-channel pole is cancelled by the kinematical zero in the Wigner function $d_{2,-1}^2(\theta)$. The  differential helicity cross section  with respect to $\cos \theta$ for the $t$-channel singularity  in the high energy limit scales like:
\beq
\frac{d\sigma^{(h_{W^+}, h_{W^-}) = (+1,-1)}}{d \cos \theta} \sim \frac{\sin^2\theta(1+\cos\theta)^2}{(1-\cos\theta)^2} \sim  \frac{(1+\cos\theta)^3}{1-\cos\theta}
\eeq
which strongly peaks in the forward region with an enhanced factor of $s/4m_t^2$. On the other hand, the anomalous top Yukawa coupling $\delta_{tth}$ appears in the longitudinal gauge bosons helicity configuration and the differential cross section in the high energy limit reads:
\beq
\frac{d\sigma^{(h_{W^+}, h_{W^-}) = (0,0)}}{d \cos \theta}  \sim \sin^2\theta
\eeq
which has its maximum near the central region $\theta \sim \pi/2$. This means that at the high energy bin, the sensitivity on the top Yukawa coupling measurement will mostly come from the central region where the transverse $W$-PDFs are suppressed.

At the threshold, the  top quark pair production from the longitudinal gauge bosons fusion is enhanced  by a factor of $m_t^4/m_W^4$. By focusing on this helicity category, the statistical significance for the top Yukawa coupling behaves as:
\beq
 \frac{S}{\sqrt{B}} \sim \sin\theta \cos\theta
 \eeq
 where for the SM background, we only include the helicity conserving top quark pair production, i.e., $(h_t, h_{\bar t}) = (\mp \frac12, \pm \frac12)$, which is a factor of 2 larger than the helicity violating ones. The significance peaks around $\theta \sim \pi/4$.

\subsection{Weak Boson PDF and Energy Scaling Behavior}
\begin{figure}[t]
\centering
\centerline{\includegraphics[width=260pt]{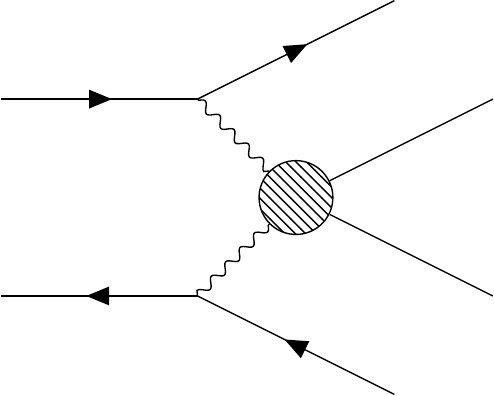}}
\caption{Illustration of the EWA approximation at the muon collider.}
\label{fig:smx}
\end{figure}

In this section, we  analyze the energy scaling behavior of $\mu^+\mu^- \rightarrow X\bar{X} \nu \bar{\nu}$ processes by making use of the Effective $W$-boson Approximation (EWA)~\cite{Dawson:1984gx,Kunszt:1987tk,Borel:2012by}. As illustrated in Fig.~\ref{fig:smx}. EWA states that at sufficiently high energies and suitable kinematical regimes, the cross section for the process $\mu^+\mu^- \rightarrow X\bar{X} \nu \bar{\nu}$ can be factorized into the on-shell hard subprocess $V \bar{V} \rightarrow X \bar{X}$ convoluted with the $W$-boson parton distribution functions:
\begin{equation}
\sigma (\mu^+ \mu^- \rightarrow X \bar{X} \nu\bar{\nu})(s) = \int_{\tau_0}^1 d\tau \sum_{ij} \Phi_{ij} (\tau, \mu_f) \hat{\sigma}(ij \rightarrow X \bar{X})(\tau s) \label{phieq}
\end{equation}
where $\sqrt{s}$  is the center-of-mass energy of muons and $\sqrt{\hat s} = \sqrt{\tau s}$ is the center-of-mass energy of the $X\bar{X}$. Here $V = W^\pm , Z$ denotes any of the SM massive electroweak gauge bosons~\footnote{We will not discuss about the $\gamma \gamma$ PDF here.}. The parton luminosity $\Phi_{ij}(\tau,\mu_f)$ is given by~\cite{Costantini:2020stv}:
\beq
 \Phi_{ij} (\tau, \mu_f) = \int_\tau^1 \frac{d\xi}{\xi} f_i(\xi,\mu_f) f_j (\frac{\tau}{\xi},\mu_f)
\eeq
Here $\mu_f$ is the factorization scale in the process under study and the weak boson PDFs at muon collider read:
\beq
\label{eq:vpdf}
\begin{split}
f_{V_\lambda} (\xi,\mu_f,\lambda=\pm 1) &= \frac{C}{16\pi^2} \frac{(g_V^\mu \mp g_A^\mu)^2 + (g_V^\mu\pm g_A^\mu)^2(1-\xi)^2}{\xi} \log\left(\frac{\mu_f^2}{M_V^2}\right)\\
f_{V_0} (\xi,\mu_f,\lambda= 0) &= \frac{C}{4\pi^2} \left((g_V^\mu)^2 + (g_A^\mu)^2\right)\left(\frac{1-\xi}{\xi} \right)
\end{split}
\eeq
The coupling constants $C, g_V^\mu, g_A^\mu$ denote the corresponding muon-weak-boson couplings and for the $W^\pm$-boson, it reads:
\beq
C = \frac{g^2}{2}, \qquad g_V^\mu = - g_A^\mu = 1
\eeq
while for the $Z$-boson, we have:
\beq
C = \frac{g^2}{\cos^2\theta_W}, \qquad g_V^\mu = \frac12 (T^3_L)^\mu + \sin^2\theta_W, \qquad g_A = -\frac12 (T^3_L)^\mu
\eeq
where we have neglected the masses of the muons. Note that $(T^3_L)^{\mu_L} = -\frac12, (T^3_L)^{\mu_R} = 0$. We will focus on the $W^+ W^-$ parton luminosity, since it is dominant compared with $ZZ$.
To obtain the energy scaling behavior of the parton luminosity $\Phi_{W^+W^-}$, we first divide the allowed values of the parameter $\tau$ into four regions: [$10^{-4}$, 0.01], [0.01,0.2], [0.2,0.8],[0,8,0.95] and then approximate the dependence  of $\Phi_{W^+W^-}$ on $\tau$ as  $\tau^{-n}$ in each region. The results are shown in Table~\ref{phitab}, where we neglected the scale-dependent logarithmic terms~\footnote{We have checked that the results won't be changed significantly by including the log terms.}. Recalling the relations  $\tau = \frac{\hat s}{s} $ and $\sqrt{\hat s} = \hat E$, the dependence on $\tau$ can be translated into the dependence on the invariant mass of $W^+W^-$ system $\hat E^{-2n}$ for constant invariant mass of $\mu^+\mu^-$ system. We can see that due to the absence of $(1-\xi)^2$ term in Eq.~(\ref{eq:vpdf}) for the plus helicity of the $W$ boson, the parton luminosity $\Phi_{W^+W^-}(\tau)$  in the $(h_{W^+}, h_{W^-}) = (+,+)$ category has the most mildest decrease as $\tau$ increases.
\begin{table}[h]
\caption{Best fit for $\Phi_{W^+W^-}$ for different ranges of $\tau$ without including the log terms. }
\label{phitab}
\centering
\tabcolsep 8pt\begin{tabular}{|c|c|c|c|c|c|}
\hline
$h_{W^+}$ & $h_{W^-}$ & $10^{-4} \leq \tau \leq 0.01$& $0.01 \leq \tau \leq 0.2$ & $0.2 \leq \tau \leq 0.8$ & $0.8 \leq \tau \leq 0.95$ \\
\hline
- & - & $\frac{1}{\tau^{1.2}}$ & $\frac{1}{\tau^{1.7}}$ & $\frac{1}{\tau^{3.9}}$ & $\frac{1}{\tau^{27}}$ \\ [8pt]
0 & 0& $\frac{1}{\tau^{1.2}}$ & $\frac{1}{\tau^{1.5}}$ & $\frac{1}{\tau^{3.0}}$ & $\frac{1}{\tau^{18}}$ \\ [8pt]
+ & + & $\frac{1}{\tau^{1.1}}$ & $\frac{1}{\tau^{1.3}}$ & $\frac{1}{\tau^{2.1}}$ & $\frac{1}{\tau^{8.0}}$ \\ [8pt]
- & + & $\frac{1}{\tau^{1.2}}$ & $\frac{1}{\tau^{1.4}}$ & $\frac{1}{\tau^{2.9}}$ & $\frac{1}{\tau^{18}}$  \\ [8pt]
+ & - & $\frac{1}{\tau^{1.2}}$ &$\frac{1}{\tau^{1.4}}$ & $\frac{1}{\tau^{2.9}}$ & $\frac{1}{\tau^{18}}$  \\ [8pt]
- & 0 & $\frac{1}{\tau^{1.2}}$ & $\frac{1}{\tau^{1.5}}$  & $\frac{1}{\tau^{3.4}}$ & $\frac{1}{\tau^{22}}$ \\ [8pt]
0 & - & $\frac{1}{\tau^{1.2}}$ & $\frac{1}{\tau^{1.5}}$ &$\frac{1}{\tau^{3.4}}$ & $\frac{1}{\tau^{22}}$ \\ [8pt]
+ & 0 & $\frac{1}{\tau^{1.1}}$&$\frac{1}{\tau^{1.4}}$  & $\frac{1}{\tau^{2.5}}$ & $\frac{1}{\tau^{13}}$ \\ [8pt]
0 & + & $\frac{1}{\tau^{1.1}}$ & $\frac{1}{\tau^{1.4}}$ & $\frac{1}{\tau^{2.5}}$ & $\frac{1}{\tau^{13}}$ \\ [8pt]
\hline
\end{tabular}
\end{table}

 Then from Eq. \eqref{phieq}, we  can see that  the differential cross section in the invariant mass of $X \bar{X}$ becomes
\begin{equation}
\frac{d\sigma}{d\hat{E}}  (\mu^+ \mu^- \rightarrow X \bar{X} \nu\bar{\nu})= \frac{ 2 \hat E}{s}  \sum_{h_2, h_2}\Phi_{W_{h_1}^+ W_{h_2}^-} (\hat E) \hat{\sigma} (W^+_{h_1}W^-_{h_2}\rightarrow X \bar{X}).
\end{equation}
Now for the most ideal scenario where the helicities of the initial and final particles can be measured and assuming that statistical error is dominant, the signal significance scales like:
\begin{equation}
\frac{S}{\sqrt{B}} \sim \frac{\frac{d\sigma_S}{d\hat{E}}}{\sqrt{\frac{d\sigma_B}{d\hat{E}}}} \sim \sqrt{\frac{\Phi_{W^+_{h_1}W^-_{h_2}}}{\hat{E}}}\mM_{\delta_{i}}^{h_1 h_2h_3h_4} \sim \frac{\mM_{\delta_{i}}^{h_1 h_2h_3h_4}}{\hat E ^{n + \frac 12}}
 \label{mumuenergyeq}
\end{equation}
where we have used the energy scaling of the parton luminosity $\Phi_{W^+ W^-} \sim \hat E ^{-2n}$ and keep  the center-of-mass energy of the muons $\sqrt{s}$ as constant. From Table~\ref{phitab}, we can see that the statistical significance decreases for the linear energy growth of BSM helicity amplitude in the whole considered regions and increases or stays constant for the quadratic energy growth for $\tau \in [10^{-4}, 0.2]$. For higher $\tau $ values ($\tau \gtrsim 0.2$),  the statistical significance decreases at least as $\hat E^{-1}$ for the quadratic energy growth of the BSM helicity amplitude. Similar conclusion holds for the fully inclusive case if we replace $\mM_{\delta_i}^{h_1 \cdots h_4}$ with $\mM_{\delta_i}^{h_1 \cdots h_4} \mM_{\rm SM}^{h_1 \cdots h_4} $, as can be seen from  the energy scaling of the statistical signal significance as follows:
\begin{equation}
\frac{S}{\sqrt{B}} \sim \frac{\frac{d\sigma_S}{d\hat{E}}}{\sqrt{\frac{d\sigma_B}{d\hat{E}}}} \sim \frac{1}{\sqrt{\hat E}}  \frac{\sum_{h_1\cdots h_4}  \Phi_{W^+_{h_1}W^-_{h_2}}  \mM^{h_1\cdots h_4}_{\rm SM}\mM^{h_1\cdots h_4}_{\delta_i}}{\sqrt{\sum_{h_1\cdots h_4}  \Phi_{W^+_{h_1}W^-_{h_2}} \left( \mM^{h_1\cdots h_4}_{\rm SM}\right)^2}}
 \label{mumuenergyeq}
\end{equation}

By using the energy scaling behavior of parton luminosity $\Phi_{W^+ W^-}$ in Table~\ref{phitab} and partonic cross section in Table~\ref{tab:xshe} for the process $\mu^+\mu^- \rightarrow t \bar{t} \nu \bar\nu$ in the presence of anomalous  couplings,  we can obtain the energy scaling for the statistical signal significance in the fully inclusive case. For the top Yukawa coupling $\delta_{tth}$ and the Higgs gauge boson coupling $\delta_{hWW}$, the result reads :
\beq 
\label{eq:sbyt}
\frac{S}{\sqrt{B}} \sim \hat E^{-1.8}, \quad \hat E^{-2.1}, \quad \hat E^{-4} , \qquad \text{for} \qquad \tau \in [10^{-4}, 0.01], \quad [0.01, 0.2], \quad [0.2, 0.8],
\eeq
where we have omitted the highest $\tau$ region. 
As expected, the sensitivity on the top Yukawa coupling decreases as bin energy  becomes larger. For the anomalous coupling $\delta \kappa_{Z, \gamma}, \delta_{Wtb}, \delta_{Zt_L}, \delta_{Zt_R}$, the sensitivity scales like:
\beq
\frac{S}{\sqrt{B}}  \sim \hat E^{0.2}, \quad \hat E^{-0.1}, \quad \hat E^{-2}, \qquad \text{for} \qquad \tau \in [10^{-4}, 0.01], \quad [0.01, 0.2], \quad [0.2, 0.8].
\eeq
from which, we can see that there is a mild increase for the signal significance at low $\tau$, a mild decrease for the intermediate $\tau$ and a decrease at high $\tau$. Finally, we find that for the anomalous coupling $\lambda_{Z,\gamma}$, the energy scaling behaves as:
\beq
\frac{S}{\sqrt{B}}  \sim \hat E^{-1.6}, \quad \hat E^{-1.7}, \quad \hat E^{-2.2}, \qquad \text{for} \qquad \tau \in [10^{-4}, 0.01], \quad [0.01, 0.2], \quad [0.2, 0.8].
\eeq
and for the coupling $\delta g_1^Z$, we have:
\beq
\frac{S}{\sqrt{B}}  \sim \hat E^{-1.6}, \quad \hat E^{-1.9}, \quad \hat E^{-3}, \qquad \text{for} \qquad \tau \in [10^{-4}, 0.01], \quad [0.01, 0.2], \quad [0.2, 0.8].
\eeq
which decreases with the energy bins.

\section{Top Yukawa couplings at the high energy muon collider}
\label{sec:yt}
In this section, we study in detail the prospects of measuring the top Yukawa coupling at a high energy muon collider. To quantify the importance of the anomalous couplings, we  parametrize the cross sections as
\begin{equation}
\sigma = \sigma_{\rm SM} \left(1+R_{1}\delta+R_{2}\delta^2\right),
 \label{req}
\end{equation}
where $\delta_i$ signifies some fractional deviation in a SM coupling. Throughout this paper, we will be primarily considering the interference term which is linear in $\delta$, but we also remark on the inclusion of the quadratic term. In terms of the kappa framework \cite{LHCHiggsCrossSectionWorkingGroup:2013rie}, $\delta_i$ and $\kappa_i$ are related by $\kappa_i = 1+\delta_i$.

\begin{figure}
\centering
\centerline{\includegraphics[width=350pt]{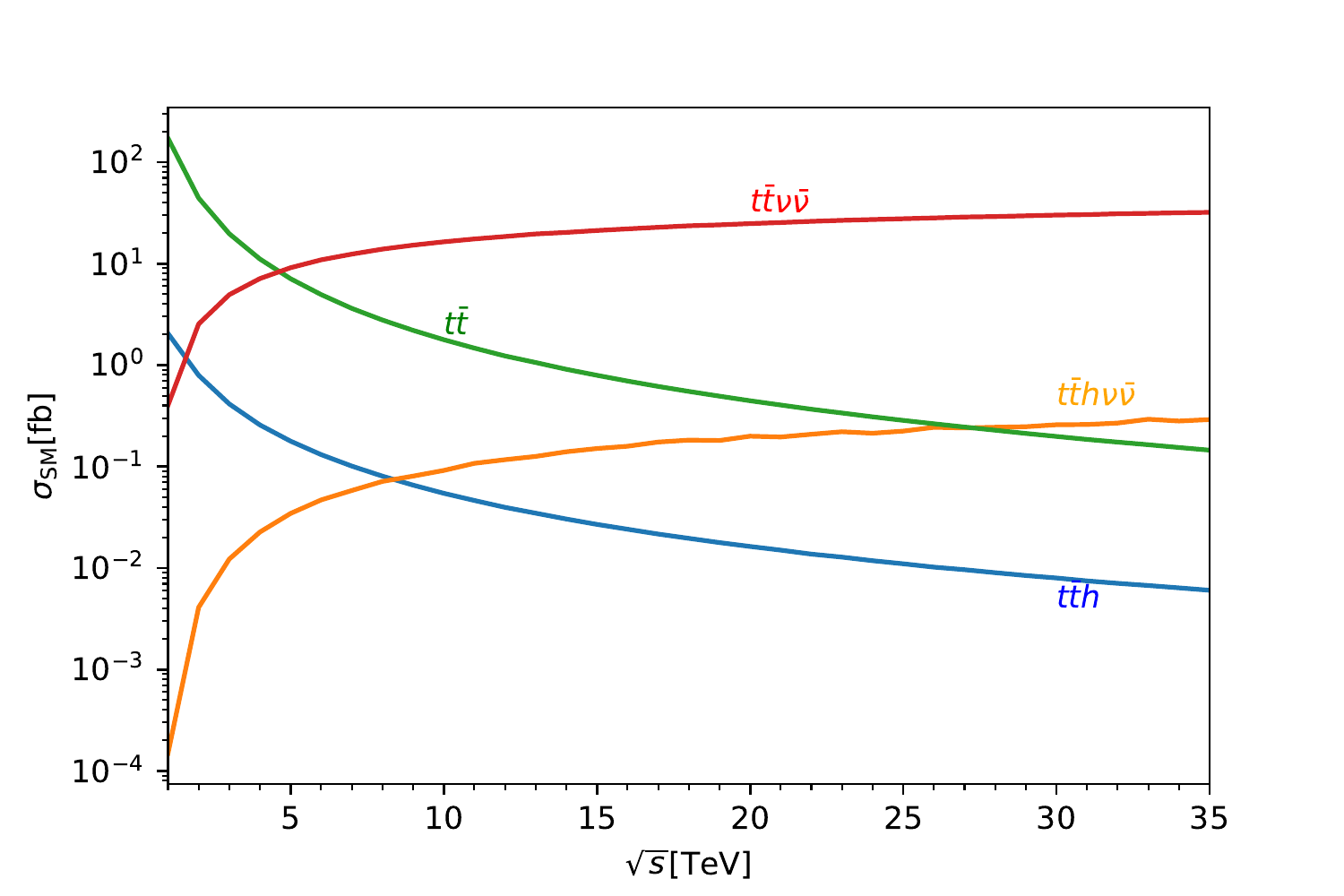}}
\caption{Cross section of SM $\mu^+ \mu^- \rightarrow t \bar{t}$, $\mu^+ \mu^- \rightarrow t \bar{t}h$, and $\mu^+ \mu^- \rightarrow t \bar{t}\nu\bar{\nu}$ with the onshell Z contribution removed.}
\label{smx}
\end{figure}

Before we present the detailed analysis for the VBF production of top quark pair, we make some comments about the Drell-Yan processes which are also involving top Yukawa coupling. The relevant processes are:
\beq
\mu^+\mu^- \rightarrow t \bar t, t \bar t h
\eeq
in which there is no energy growing behavior for the anomalous top Yukawa coupling $\delta_{tth}$. In Fig.~\ref{smx}, we have plotted the SM cross sections as functions of center-of-mass energy of the muon collider for both DY and VBF productions of  top quark pair and top quark pair plus a Higgs boson. We can see that due to the logarithmic 
growth of the VBF processes and the inverse of energy squared decrease of the DY processes, the VBF productions start to become dominant at 5 (8) TeV center-of-mass energy for the $t\bar t (t\bar t h)$. Besides the small cross sections at the  high energy muon collider, the $R$-values defined in in Eq.~(\ref{req}) are also very small for the DY production of top quark pair.  In order to have the $t\bar t$ process  involve the top Yukawa coupling, it is necessary to include the non-zero muon masses. In this case,  the dependence of the cross section on $\delta_{tth}$ will be suppressed by the muon Yukawa coupling squared $m_\mu^2/v^2 \sim 2 \times 10^{-7}$. We have checked that for this process, the $R$-ratios defined in Eq.~(\ref{req}) for the anomalous coupling $\delta_{tth}$ are very small:
\beq
\begin{split}
R_1=2.337\times 10^{-5}, \qquad R_2 = 1.169 \times 10^{-5} \qquad  @ \text{10 TeV} \\
R_1=2.343\times 10^{-5}, \qquad R_2 = 1.172 \times 10^{-5}\qquad @\text{30 TeV} \\
\end{split}
\eeq
and we will not consider it any further.  For the  DY process $\mu^+ \mu^- \rightarrow t\bar t h$, the $R$-values are:
\beq
\begin{split}
R_1= 1.62, \qquad R_2 =  0.797 \qquad \text{at 10 TeV} \\
R_1=1.56 \qquad R_2 = 0.774\qquad \text{at 30 TeV} \\
\end{split}
\eeq
We can see that the $R$-values stay almost constant as the center-of-mass energy of the muon collider increases. We expect that the sensitivity on the top Yukawa coupling from this process will come from the lower energy stages of the muon collider. Such analysis has been performed at CLIC in the baseline energy of 1.4 TeV~\cite{CLICdp:2018esa}.

\subsection{Simulation and Cuts}

We now turn to the simulation and analysis of the process $\mu^+\mu^- \rightarrow t \bar t \nu_\mu  \bar\nu_\mu, t\bar t h \nu_\mu \bar{\nu}_\mu$ in the presence of the anomalous top Yukawa coupling $\delta_{tth}$.   We are using {\tt Madgraph5}  \cite{madgraph} to calculate the cross sections and generate the events at LO. The anomalous coupling $\delta_{tth}$ is implemented by using the BSMC model file~\cite{bsmc}. We will work at the level of top quarks and no decaying of the top quarks will be simulated.

One advantage of the lepton colliders compared with hadron collider is that the initial energies of the colliding  leptons are known very precisely~\cite{ALEPH:2005ab}, as a result, the invariant mass of the two outgoing neutrinos is indirectly determined by the momenta of the top quark pair or the top quark pair plus Higgs boson. This is defined as recoil mass and for the $t \bar t \nu_\mu \bar \nu_\mu$ process,
\beq
M_{\rm recoil}^2 = (p_{\mu^+} + p_{\mu^-} - p_{t} - p_{\bar t})^2,
\eeq
For the $t \bar t h \nu_\mu \bar \nu_\mu$ process, it is given by:
\beq
M_{\rm recoil}^2 = (p_{\mu^+} + p_{\mu^-} - p_{t} - p_{\bar t}- p_h)^2.
\eeq
We will impose the following cut  on the recoil mass at the generator level:
\beq
M_{\rm recoil} > 200 \text{GeV},
\label{eq:recoil}
\eeq
which will remove the contribution from the process $ t \bar t Z \rightarrow t \bar t (\nu \bar \nu)$.  In  Table \ref{background}, we have presented the cross sections of the VBF $t\bar t$ production and the potential relevant backgrounds for some benchmark scenarios at the high energy muon collider.  For  all the VBF processes, the cross sections are presented after the cut in Eq.~(\ref{eq:recoil}).

\begin{table}[t]
\center
\caption{Cross sections for signal and background. For the VBF processes, the cut on the recoil mass in Eq.~(\ref{eq:recoil}) has been imposed.}
\label{background}
\tabcolsep10pt\begin{tabular}{|c|c|c|c|c|c|}
\hline
\textbf{$\sqrt{s}$ (TeV)} \textbackslash  \textbf{$\sigma_{\rm SM}$(fb)} & 3 & 6 & 10 & 14 & 30  \\
\hline
$ t\bar{t} \nu_\mu\bar{\nu}_\mu $& 4.93 & 10.9 & 16.4 & 20.5 & 30.1\\ [7pt] 
$t \bar{t} h \nu_\mu \bar{\nu}_\mu$& 0.0121 & 0.0460 & 0.0914 & 0.141 & 0.269 \\ [10pt] 
 \hline
$t \bar{t}$ & 19.7 & 4.95 & 1.78 & 0.909 & 0.198 \\ [10pt]  
$t \bar{t} h$ &  0.414 & 0.131 & 0.0547 & 0.0305 & 0.00793 \\ [10pt]
$W^+W^- \nu_\mu \bar{\nu}_\mu $ & 120 & 259 & 399 & 515 & 815  \\ [10pt]
$W^\pm Z \mu^\mp (\bar{\nu}_\mu/ \nu_\mu)$  \tablefootnote{Sum of the cross sections for $W^+ Z \mu^- \bar{\nu}$ and $W^- Z \mu^+ \nu$ with $p_T> 30$ GeV for charged leptons and the on-shell $W \rightarrow \mu \nu$ contribution removed.} & 96.6  & 215 & 340 & 443  & 717 \\ [10pt]
\hline
\end{tabular}
\end{table}

\begin{figure}[t]
\centering
\subfloat[Standard Model $\theta_t$. ]{\includegraphics[width=0.5\textwidth]{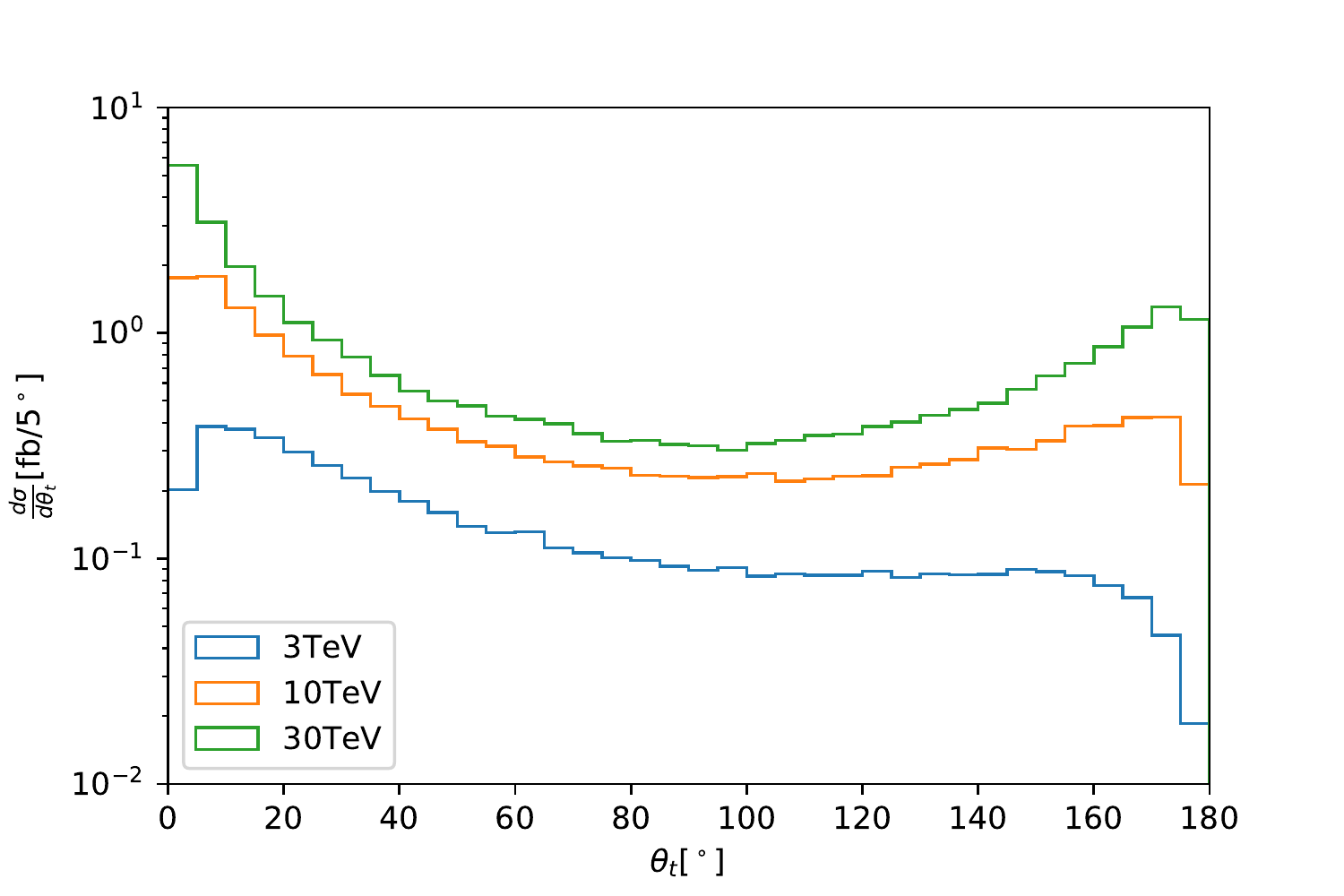}}
\subfloat[Standard Model $\slashed{E}_T$ of the top]{\includegraphics[width=0.5\textwidth]{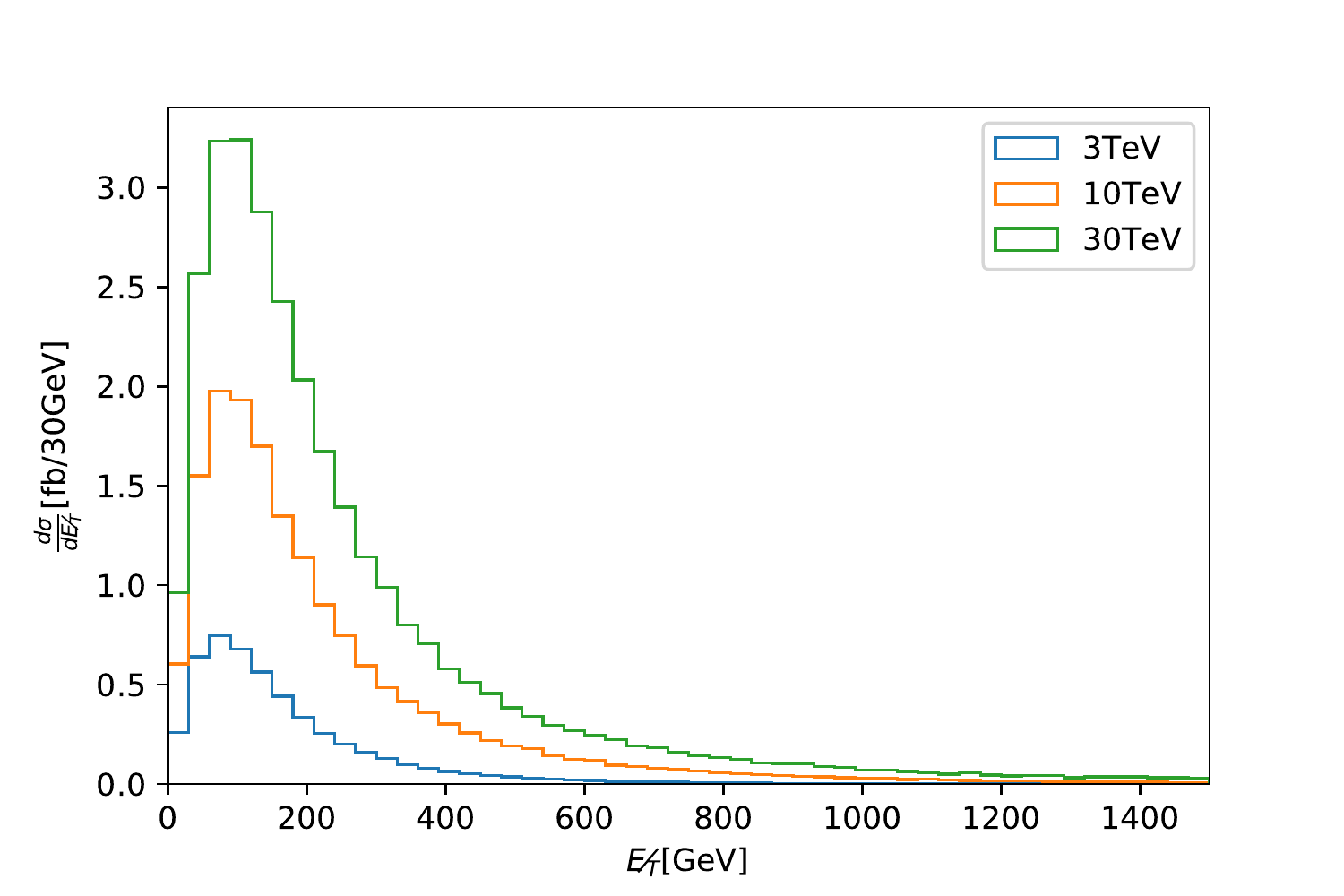}}
\caption{Standard Model distribution of $\theta$ and $p_T$ of the top  quark at 3, 10, and 30 TeV muon colliders after the cut on the  recoil mass in Eq.~(\ref{eq:recoil}).}
\label{smdiff}
\end{figure}

The decaying branching ratios for the top quark pair are respectively
45\%, 28\%, 4.4\% in the fully hadronically decaying channel,  semi-leptonically decaying channel and fully leptonically decaying channel \cite{ParticleDataGroup:2020ssz} \cite{Workman:2022ynf}~\footnote{In the estimation of the decaying branching ratios, we have neglected the $\tau \nu$ decay of the $W$ bosons. Including it will have mild effects on the final results.}. We will focus on the semi-leptonically decaying channel where the top quark and anti-top quark can be reconstructed and distinguished by the charges of the decayed leptons. To suppress the beam induced background, we put the following cuts on the polar angles of the top quark pair in the laboratory frame:
\beq
10^\circ <\theta_{t, \bar t} < 170^\circ
\label{eq:thetacut}
\eeq
where in our convention, the $z$-axis align with the direction of the $\mu^+$ beam. As shown in Fig.~\ref{smdiff}, the $\theta_t$ distribution peaks  strongly in the forward region  at 3, 10, 30 TeV muon collider and  peaks also mildly in the backward region for 10, 30 TeV center-of-mass energy. 
The cut efficiencies for the $\theta_{t, \bar{t}}$ cuts at the 10 TeV and 30 TeV muon collider are 0.57 and 0.43, respectively.
This reduces the cross sections of the SM $t\bar{t} v\bar v$ in the semi-leptonically decaying channel to 2.63 fb and 3.61 fb for 10 TeV and 30 TeV muon collider respectively. Here the numbers have also taken into account the  branching ratios of the semi-leptonically decaying channel of top quark pair. 

We expect that the signal manifests itself in the kinematical region where effective $W$ approximation applies as this is the hard scattering regime. To maximize the sensitivity and also to help to reconstruct the effective $W$ boson partonic center-of-mass frame, we impose the following criterion:
\beq
\slashed E_T < 200 \text{GeV}
\label{eq:etcut1}
\eeq
where at the truth-level, the missing transverse energy $\slashed E_T$ is equal to the magnitude of the transverse momentum of the two neutrino system or top quark pair system:
\beq
\slashed E_T = |\vec p_{T, \nu} +\vec p_{T, \bar\nu}  | =  |\vec p_{t, t} +\vec p_{T, \bar t}  |.
\eeq
Note that we also require the missing transverse energy to be larger than 20 GeV
\beq
\slashed E_T  > 20 \text{GeV},
\label{eq:etcut2}
\eeq
which is used to reduce the background from DY production of $t\bar t$ with initial state radiation or bremsstrahlung effects~\cite{CLICdp:2018esa}.
The cut efficiencies we obtain from comparing the $\slashed E_T$ and  $\theta_{t, \bar{t}}$ cuts to the  $\theta_{t, \bar{t}}$ cuts alone are 0.50 and 0.44 for 10 TeV and 30 TeV, which further reduces the SM cross sections to 1.32 fb and 1.59 fb, where again we include the semi-leptonic branching ratio. This sizable suppression from $\slashed E_T$ cut is as expected as from Fig.~\ref{smdiff}.  For illustration, in Table~\ref{ttrval}, we have listed the values of the SM cross sections in the semi-leptonically decaying channel and the $R_{1,2}$ in different bins of $m_{t\bar t}$ for the VBF production of  top quark pair  after all the preliminary  cuts in Eq.~(\ref{eq:recoil}), (\ref{eq:thetacut}), (\ref{eq:etcut1}),(\ref{eq:etcut2}) at 30 TeV muon collider. We can see that there is no energy growing behavior for the interference term, as expected from the previous  analytical study. On the other hand, we do see the $R$-value for the squared term possess larger values at higher energy bins. For comparison, we have also presented the SM cross sections  $R$-values for the process $\mu^+ \mu^- \rightarrow t \bar{t} h \nu \bar{\nu}$ with semi-leptonically decaying top quark pair and Higgs decaying to bottom quark pair at 30 TeV muon collider. We can see that there is indeed energy growing behavior for the linear term. 

\begin{table}[H]
\centering
\caption{The SM cross sections and the $R$-values for anomalous  top Yukawa coupling in the process $\mu^+ \mu^- \rightarrow t \bar{t} \nu \bar{\nu}, t\bar t h \nu \bar \nu$ after all the preliminary cuts  in Eq.~(\ref{eq:recoil}), (\ref{eq:thetacut}), (\ref{eq:etcut1}), (\ref{eq:etcut2}) with semi-leptonic decay for $t\bar{t}$, and $b\bar{b}$ decay for the Higgs boson in different invariant mass bins at 30 TeV muon collider.} 
\label{ttrval}
\tabcolsep 10pt\begin{tabular}{|c|c|c|c|c|c|}
\hline
$m(\bar{t}t)$  & $\sigma_{\rm SM}$ (fb)  & $R_1$ & $R_2$ \\
\hline
0-1TeV & $1.28 $  & -0.0803  & 1.33   \\
1-5TeV & $0.325$ & -0.220 & 12.3 \\
5-10TeV& $0.00538$  & -0.155 & 157 \\
10-15TeV& $4.17 \cdot 10^{-4} $ & -0.152 & 468  \\
15-20TeV & $5.21 \cdot 10^{-5}$  & -0.163  &886 \\
20-25TeV& $6.36 \cdot 10^{-6} $ & -0.0608  & 1199 \\
25-30TeV& $1.06 \cdot 10^{-6} $ & -0.00202   & 355 \\
\hline

\end{tabular} \\
\vspace{0.5cm}
\begin{tabular}{|c|c|c|c|c|c|c|c|}
\hline
$m(\bar{t}th)$& $\sigma_{\rm SM}$ (fb) & $R_1$ & $R_2$ \\
\hline
0-1TeV& $1.10 \cdot 10^{-3}$ & 5.75 & $15.5 $  \\
1-5TeV & $2.74 \cdot 10^{-3} $& 7.73 & $320 $ \\
5-10TeV & $1.72 \cdot 10^{-4}$ & 26.8  &$9090 $  \\
10-15TeV & $2.14 \cdot 10^{-5} $ & 49.8 & $51400 $\\
15-20TeV & $3.48 \cdot 10^{-6} $ & 72.8  & $147000$ \\
20-25TeV  & $7.44 \cdot 10^{-7}$ & 58.7   & $186000 $\\
25-30TeV& $1.16 \cdot 10^{-7} $ & 16.5  & $76500$ \\
\hline
\end{tabular}
\end{table}

\begin{figure}[th]
\centering
\subfloat[$\theta^*$ (the angle between the top and the $W^+$ in the $W^+ W^-$ center of mass) distribution]{\includegraphics[width=0.5\textwidth]{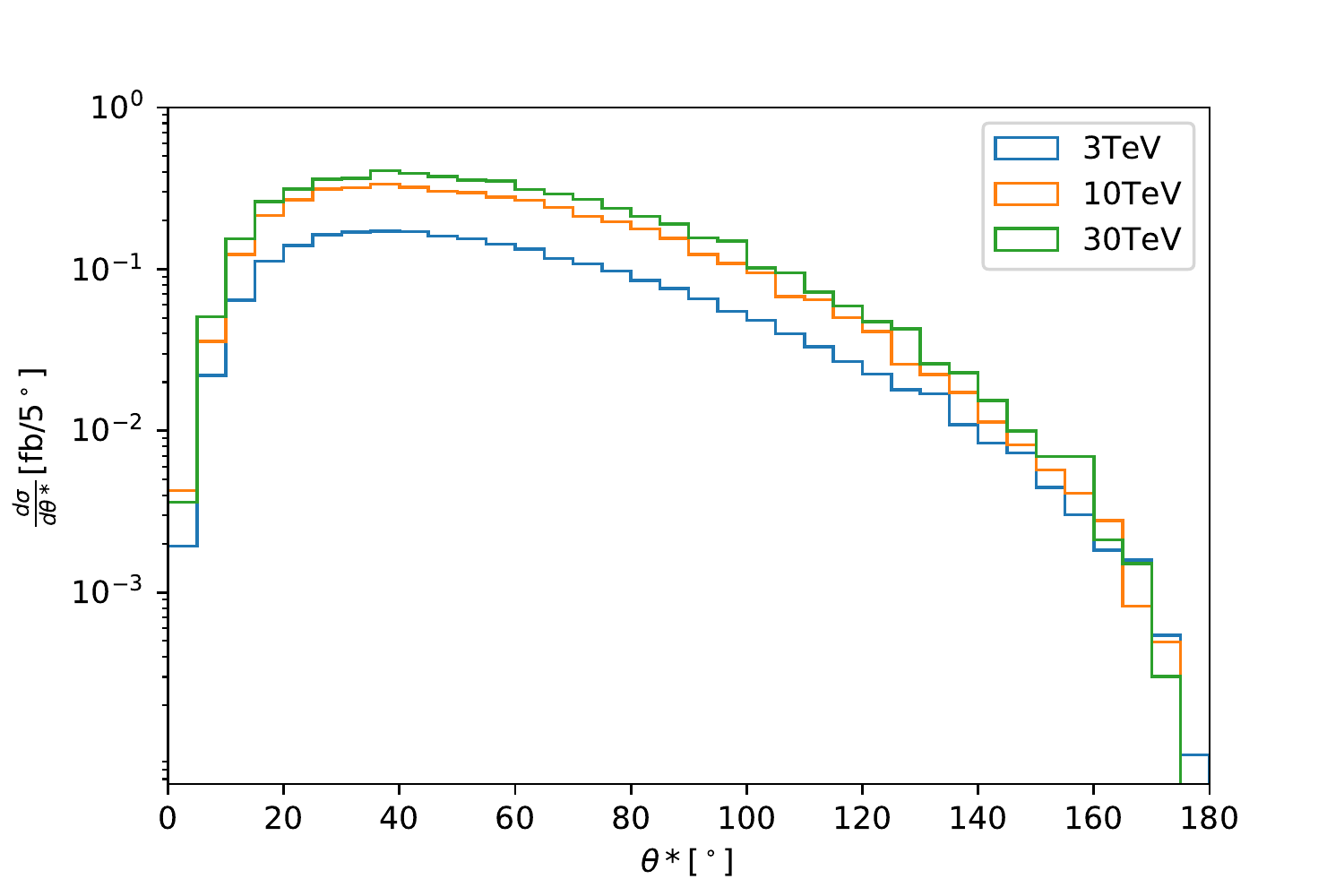}}
\subfloat[$\theta^*$ (the angle between the top and the $W^+$ in the $W^+ W^-$ center of mass) distribution. ]{\includegraphics[width=0.5\textwidth]{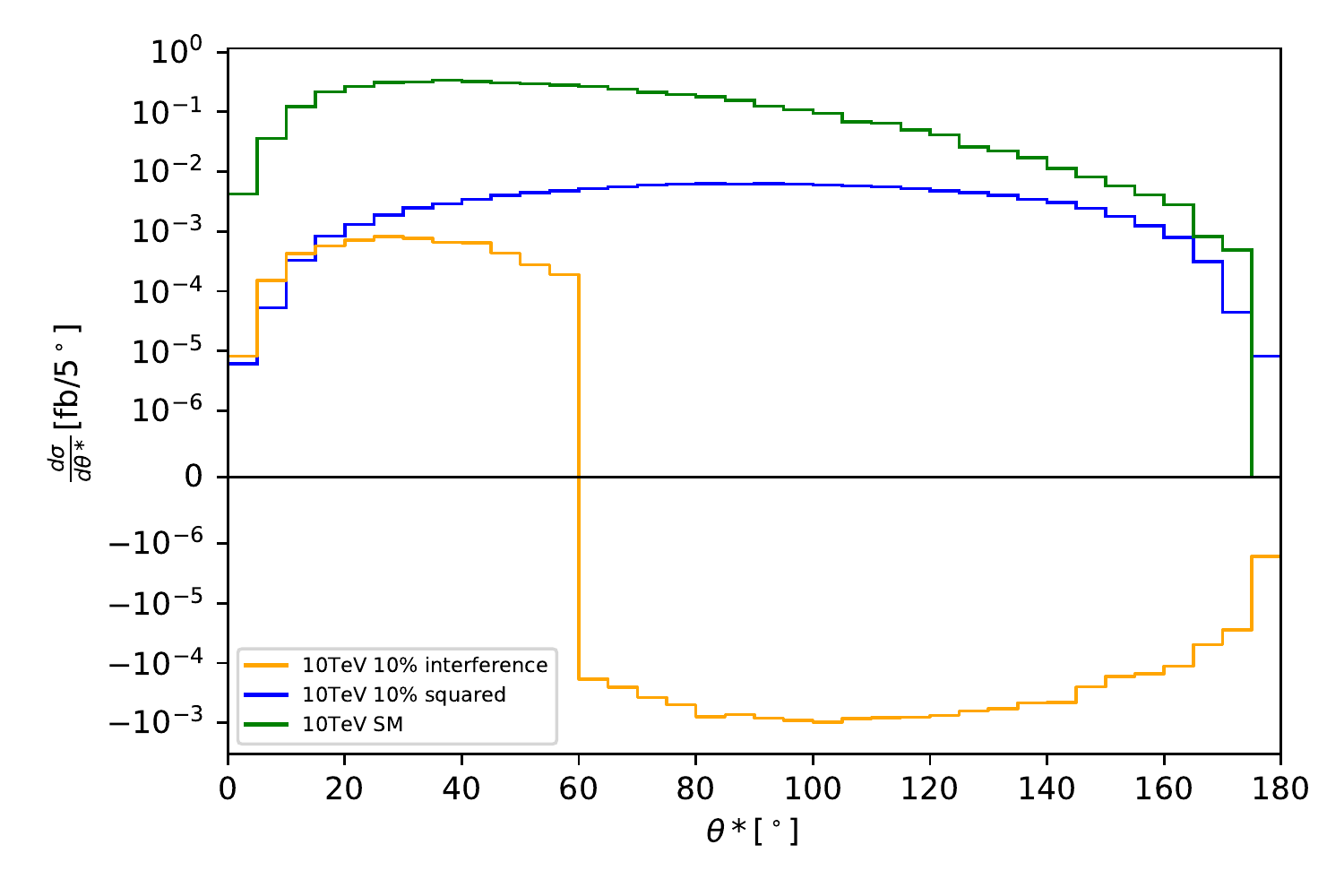}}
\caption{ The distributions of $\theta^*$  for the SM (left panel) and $\delta_{tth} = 10\%$ (right panel) after the all the preliminary cuts in Eq.~(\ref{eq:recoil}), (\ref{eq:thetacut}), (\ref{eq:etcut1}), (\ref{eq:etcut2}).}
\label{smdiff2}
\end{figure}

As discussed in previous sections and also shown in Fig.~\ref{smdiff2}, the scattering angle  in the partonic center-of-mass frame $\theta^*$ can be used to enhance the sensitivity to the top Yukawa coupling. Here we have used an asterisk to distinguish between the polar angle of top quark in the $W^+ W^-$ frame and the polar angle in the $\mu^+\mu^-$ frame. Furthermore, in determining the scattering angle $\theta^*$ in the partonic frame, we assume that the neutrinos are collinear with the muon beams. To be explicit,  the scattering angle $\theta^*$ can be obtained from the kinematical variables in the lab frame as follows:
\begin{equation}
\tan \theta^* = \frac{\sqrt{p_{t,x}^2+p_{t,y}^2}m_{t\bar{t}}}{-E_{t}\, \vec{p}_{t\bar t}+ p_{t,z} \, E_{t\bar t}}. \label{thetastareq}
\end{equation}
where $p_{t,x}$ is the $x$-component of the momentum of the top quark and similarly for the $p_{t,y}, p_{t,z}$. $m_{t\bar t}$ is the invariant mass of the top quark pair and $(E_{t\bar t}, \vec{p}_{t\bar t})$ is the four-momentum of the top quark pair. Here we have used the fact that the transverse momentum of the top quark is the same in both frame and the $z$-component of the momentum of the top quark in the partonic frame is obtained by a boost.

\begin{table}[H]
\small
\center
\caption{Efficiencies from CLIC analysis of the semi-leptonically decaying channel with $\text{P}(e^-) = -80\%$~\cite{CLICdp:2018esa}.}
\label{tab:eff}
\tabcolsep10pt\begin{tabular}{|c|c|c|c|c|c|}
\hline
\textbf{$\sqrt{s}$} & 380 GeV & 1.4 TeV ($\sqrt{s^\prime} \geq 1.2$ TeV)& 3 TeV $(\sqrt{s^\prime} \geq 2.6$ TeV)  \\
\hline
$\epsilon_{\rm eff} (e^+ e^- \rightarrow t \bar t \rightarrow qqqq l\nu ) $ & 64\%& 37\%& 33\% \\
\hline
\end{tabular}
\end{table}

In addition to the invariant mass bins of the top quark pair in Table~\ref{ttrval}, we also divide the scattering angle $\theta^*$ into six bins with bin width of $30^\circ$. The corresponding cross sections and $R$-values in each two-dimensional bin are shown in Table~\ref{tab:xs10}, ~\ref{tab:xs30} and Table~\ref{tab:rval10}, ~\ref{tab:rval30} respectively in Appendix~\ref{app:errorxs}. In order to take into account the  reconstruction efficiencies of the semi-leptonically decaying top quark pair, we have extracted the numbers from the analysis of top quark pair production at 380 GeV, 1.4 TeV and 3 TeV center-of-mass energy of CLIC~\cite{CLICdp:2018esa}. The results are listed in Table~\ref{tab:eff}. We will use the following values for the reconstruction efficiencies for different $m_{t\bar t}$ bins:
\beq
[0, 1] \text{TeV}: \quad 64 \%, \qquad \text{all other bins}: \quad   33\%
\eeq
and assume that the  SM reducible backgrounds has been reduced to a negligible level.
Similar efficiencies apply to the bins of $m_{t\bar t h}$ for the process  $\mu^+ \mu^- \rightarrow t \bar{t} h \nu \bar{\nu}$ with the Higgs boson decaying into bottom quark pair $h\rightarrow b \bar b$ with a branching ratio of $58\%$ \cite{ParticleDataGroup:2020ssz} \cite{Workman:2022ynf}.

\section{Results and Discussion}
\label{sec:result}
\begin{figure}[t]
\centering
\subfloat[$\mu^+ \mu^- \rightarrow t \bar{t} \nu \bar{\nu}$ with $\sqrt{s} = 10$ TeV\\ and $L = 10 \, \text{ab}^{-1}$.]{\includegraphics[width=0.5\textwidth]{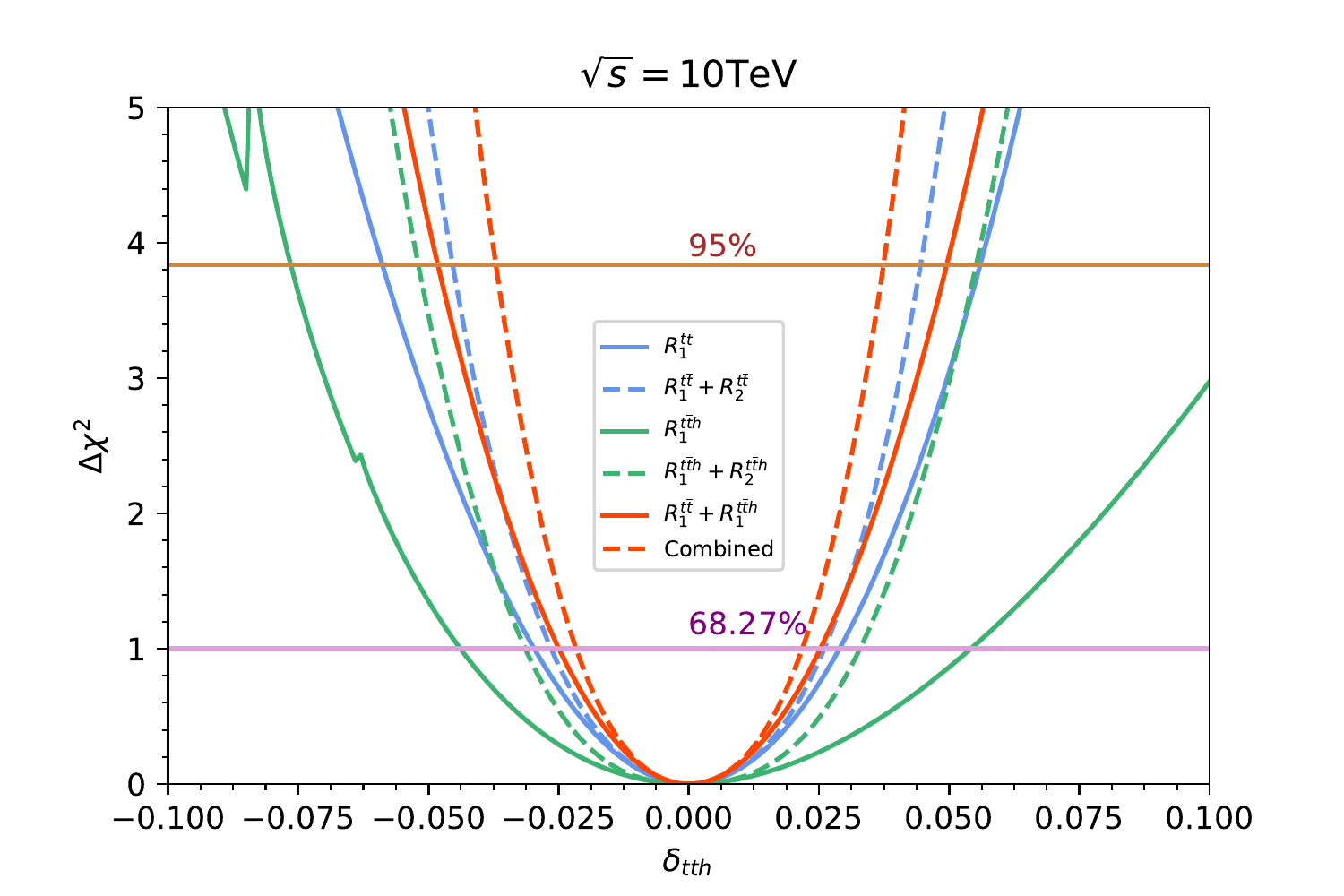}}
\subfloat[ $\mu^+ \mu^- \rightarrow t \bar{t} \nu \bar{\nu}$ with $\sqrt{s} = 30$ TeV \\ and $L = 90 \, \text{ab}^{-1}$.]{\includegraphics[width=0.5\textwidth]{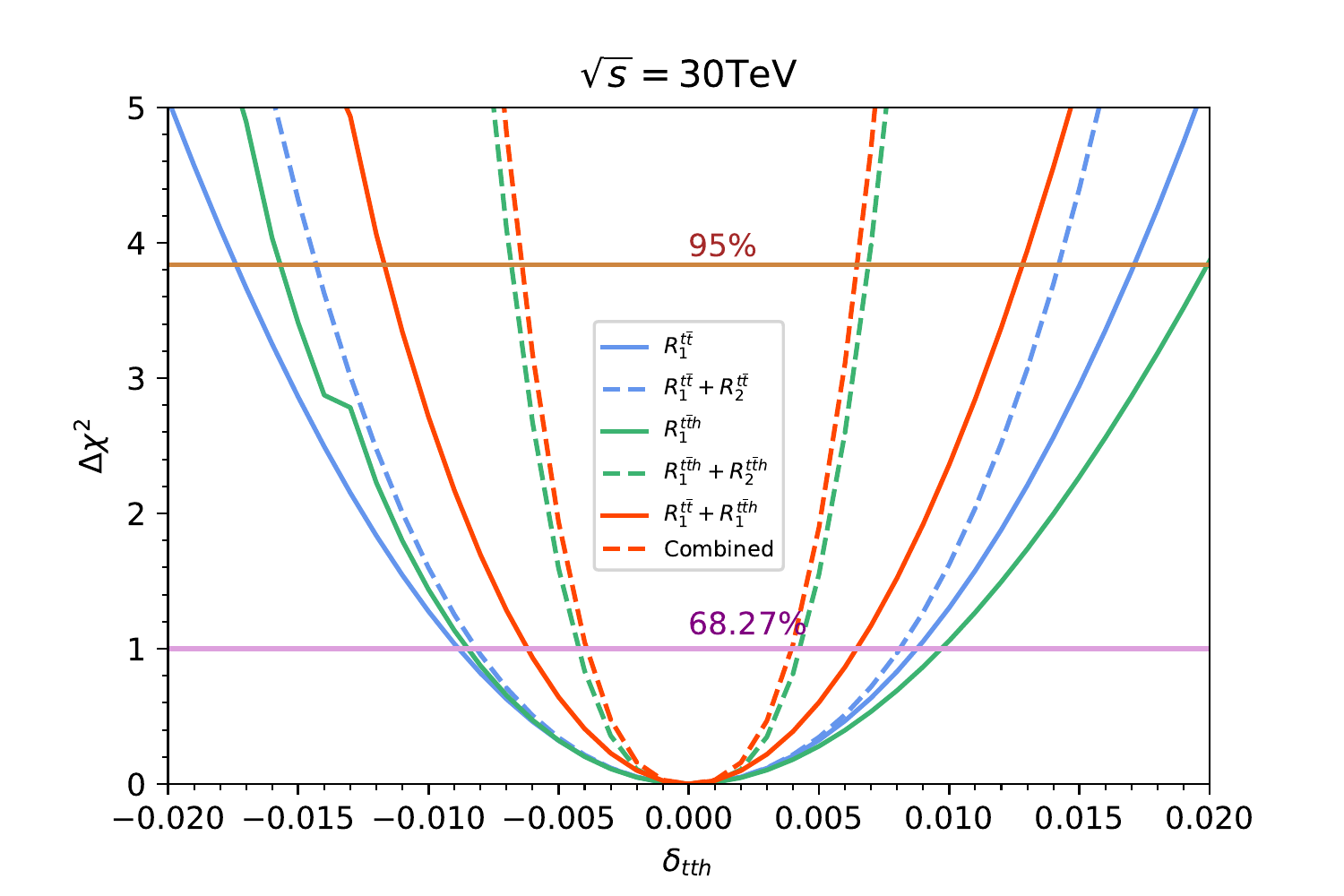}}
\caption{$\Delta \chi^2$ plot as a function of anomalous top Yukawa coupling $\delta_{tth}$ for processes $\mu^+\mu^- \rightarrow t \bar{t} \nu \bar{\nu}$ and $\mu^+\mu^- \rightarrow t \bar{t} \nu \bar{\nu}h$ at 10 TeV (left panel) and 30 TeV (right panel) muon collider. Here $R_1 (R_2)$ denotes the interference term and  the squared term respectively.}
\label{rvalsplot}
\end{figure}
We follow the procedure in Appendix~\ref{app:stat} to construct the likelihood functions by combing all the two dimensional bins defined in Table~\ref{tab:xs10}  and Table~\ref{tab:xs30} for 10 TeV, 30 TeV muon collider correspondingly. The integrated luminosity is assumed to be 10  (90)$\text{ab}^{-1}$ at 10 (30)TeV muon collider. The $\Delta \chi^2$ as functions of the anomalous top Yukawa coupling $\delta_{tth}$ for the semileptonically  decaying channels of the $t\bar t \nu\bar \nu , t \bar{t} h \nu \bar \nu$ are presented in Fig.~\ref{rvalsplot}. For each process, we have considered two cases: with only the linear term $R_1$ and with both the linear term $R_1$ and the quadratic term $R_2$. The $95\%$ C.L. interval for the $\delta_{tth}$ for different scenarios are shown in Table~\ref{constraints}. We find that due to the lack of energy growing behaviors in the $t\bar t \nu \bar \nu$, the expected sensitivity on the anomalous top Yukawa coupling $\delta_{tth}$ is not majorly affected by the inclusion of the quadratic term at both 10 TeV and 30 TeV. In contrast, for the $t\bar t h \nu \bar \nu$, the quadratic terms can make a big difference (a factor of 2-3) on the top Yukawa coupling sensitivity, which is a reflection of the energy growing effects. For this process, a dedicated study should be provided to address the issue of the effective field theory breaking down, which we leave for future work. Here we are focusing on the results obtained by including the linear term $R_1$ only. At 10 (30) TeV muon collider, the 95\% C.L. on the anomalous coupling $\delta_{tth}$  from $t\bar t \nu \bar \nu$ reads 5.6\% (1.7\%), which is generally in agreement with the results of~\cite{AlAli:2021let}. These can be compared with 4\% and 2 \% projections at 95\% C.L.  for the HE-LHC under the base and optimal scenarios respectively \cite{Cepeda:2019klc} as well as the 2\% projection at a 100TeV collider \cite{Mangano:2015aow}, which are also listed in Table~\ref{constraints}. For the process $t\bar t h \nu \bar \nu$, without worrying about the issues of EFT mentioned earlier, we find that the result is comparable with $t \bar t \nu \bar \nu$, especially at 30 TeV muon collider. It deserves further detailed study, which we leave for future work.



\begin{table}[H]
\centering
\caption{$95\%$ C.L. on the  anomalous top Yukawa coupling $\delta_{tth}$ for different scenarios at 10 TeV and 30 TeV muon collider. } 
\label{constraints}
\tabcolsep 10pt\begin{tabular}{|c|c|c|}
\hline
$\sqrt{s_{\mu^+\mu^-}}$ & Process & Sensitivity \\
\hline
\multirow{6}{*}{10 TeV @ 10 ab$^{-1}$ } & $t\bar{t}\nu\bar{\nu} $ \  $R_1$ & $[-5.9\%, 5.6\%]$  \\
 & $t\bar{t}\nu\bar{\nu} $ \ $R_1 + R_2$ & $[-4.5\%, 4.5\%]$  \\
 &$t\bar{t}h\nu\bar{\nu}$   \  $R_1$  & $[-7.6\%, 12\%]$ \\
  & $t\bar{t} h \nu\bar{\nu} $ \ $R_1 + R_2$ &  $[-5.2\%, 5.5\%]$  \\
 & $t\bar{t}\nu\bar{\nu} + t\bar{t}h\nu\bar{\nu}$   \  $R_1$ & $[-4.8\%, 5.0\%]$  \\
  & $t\bar{t}\nu\bar{\nu} + t\bar{t}h\nu\bar{\nu}$   \  $R_1 + R_2$ & $[-3.7\%, 3.7\%]$  \\
\hline
\multirow{6}{*}{30 TeV @ 90 ab$^{-1}$}  & $t\bar{t}\nu\bar{\nu}$ \ $R_1$ & $[-1.7\%, 1.7\%]$  \\
 & $t\bar{t}\nu\bar{\nu}$ \ $R_1 + R_2$& $[-1.4\%, 1.4\%]$  \\
 & $t\bar{t}h\nu\bar{\nu}$ \ $R_1$ & $[-1.6\%, 2.0\%]$  \\
& $t\bar{t} h \nu\bar{\nu} $ \ $R_1 + R_2$ &   $[-0.68\%, 0.69\%]$ \\
 & $t\bar{t}\nu\bar{\nu} + t\bar{t}h\nu\bar{\nu}$   \  $R_1$ & $[-1.2\%, 1.3\%]$  \\
 & $t\bar{t}\nu\bar{\nu} + t\bar{t}h\nu\bar{\nu}$   \  $R_1 + R_2$ & $[-0.64\%, 0.65\%]$  \\
\hline
\multicolumn{3}{|c|}{Other Colliders}\\
\hline
\multirow{1}{*}{14 TeV HL-LHC @ 3 ab$^{-1}$} & $t \bar t h \rightarrow \text{Multiple Leptons}$ &6.9\%~\cite{Cepeda:2019klc} \\

\hline 
1.4 TeV CLIC @ 1.5 ab$^{-1}$ & $t\bar{t}h \rightarrow  6 j  + b \bar{b}, \ell \nu 4 j + b \bar{b}$ & 7.4\%~\cite{CLICdp:2018esa}\\
\hline 
100 TeV Collider @ 20 ab$^{-1}$ & $t\bar{t}h \rightarrow \ell \nu 4 j + b \bar{b}$ & 2\%~\cite{Mangano:2015aow}\\
\hline
\end{tabular}
\end{table}

\section{Conclusion}
\label{sec:conclu}

In this paper, we have performed a detailed analysis about the measurement of the top Yukawa coupling at the high energy muon collider by studying the  process $\mu^+\mu^- \rightarrow t \bar{t} \nu\bar\nu$. In particular, we have studied  the energy scaling behavior of statistical  signal significance $S/\sqrt{B}$ for the subprocess $W^+W^- \rightarrow t \bar t$ and for the full processes at the muon collider by employing the effective $W$-boson approximation. In addition, we have presented the explicit formulae for the helicity amplitudes for the subprocess $W^+W^- \rightarrow t \bar t$ in the presence of anomalous couplings, where for completeness, we have also included anomalous triple gauge boson couplings and anomalous gauge-boson-fermions couplings. The high energy limits of the different helicity amplitudes are shown in Table~\ref{tab:hche1}, \ref{tab:hche2}, whereas the threshold behaviors are given in Table~\ref{tab:hcth1}, \ref{tab:hcth2}. We have found that the sensitivity on the anomalous top Yukawa coupling $\delta_{tth}$ decreases as the energy of the bin increases as shown in Eq.~(\ref{eq:sbyt}). This is partially  due to the fact that the SM amplitude for the helicity configurations $(0,0,\pm \frac 12, \pm \frac 12)$ scales like $m_t/\hat E$. As a result,  the interference  between the SM and  BSM amplitudes  will stay constant instead of growing linearly with $\hat{E}$. Secondly, the suppression of the parton luminosity $\Phi_{W^+W^-}(\tau)$ at high $\tau $ also reduces the signal significance $S/\sqrt{B}$ at high energy bins. As a byproduct, we also found that in the case of triple-gauge- boson couplings $\delta \kappa_{Z,\gamma}$ and the gauge-boson-fermion couplings $\delta_{Wtb}, \delta_{Zu_L}, \delta_{Zu_R}$, the statistical signal significance mildly increases for small values of $\tau$, mildly decreases for intermediate values of  $\tau$, and decreases at large $\tau$ values. 

The semi-analytical analysis has been confirmed by our  numerical simulation,  where we studied  the prospects on the top Yukawa coupling measurement at 10 TeV and 30 TeV muon colliders. We have imposed the basic selections cuts  in Eq.~(\ref{eq:recoil}), (\ref{eq:thetacut}), (\ref{eq:etcut1}), (\ref{eq:etcut2}) and focused on the semi-leptonically decaying channel of the top quark pair. The reconstruction efficiencies in this channel have been extracted from the CLIC analysis for different stages. Similar efficiencies are also  applied  to the $t \bar t h \nu_\mu \bar \nu_\mu$ process, where the Higgs boson is assumed to decay into a bottom quark pair. Furthermore, we used the  distribution  of the scattering angle in the partonic center-of-mass frame for the $t \bar t \nu_\mu \bar \nu _\mu$ to enhance sensitivity.  The  precision on the  anomalous top Yukawa coupling at the 95 \% C.L. is projected to be 5.6\% (1.7\%)  for VBF production of a top quark pair at a 10 (30) TeV muon collider. The precision  from VBF production of $t\bar t h$ is comparable to the top quark pair, but is sensitive to contributions from the quadratic term. Therefore, it demands further detailed study, which we leave for future possible work. 

\section*{Acknowledgments}

We would like to thank Markus Luty for the collaboration in the early stage of the project. The work of MC and DL was supported by DOE Grant Number DE-SC-0009999.

\appendix

\section{Cross sections,  the $R$-values and errors}
\label{app:errorxs}
In this appendix,  we list the cross sections and the $R$-values for the two-dimensional bins in terms of  $m_{t\bar t}, \theta^*$. The cross sections are presented in Table~\ref{tab:xs10} for 10 TeV muon collider and in Table~\ref{tab:xs30} for 30 TeV muon collider. The $R$-values are given in Table~\ref{tab:rval10} for 10 TeV muon collider and in Table~\ref{tab:rval30} for 30 TeV muon collider. The errors in the tables are associated with the limited number of events generated by {\tt Madgraph5}~\cite{Alwall:2014hca} and we describe about how to obtain them in the following. Note that we do not take into account the errors when we make the projections for the top Yukawa coupling measurement.

The cross section of a given process for some set of cuts is
\begin{equation}
\sigma = \frac{\sum_i w_i}{N}
\end{equation}
where the $w_i$ are the weights of events that remain after the cuts and $N$ is the total amount of events in the run~\footnote{In order to find cross sections across multiple LHE files with different cuts, we simply sum the individual over cross sections.}. In the case where all events have positive weight, the error is the familiar $\frac{1}{\sqrt{N}}$. However, the error increases when roughly half of the events have negative weight.

To determine the error in the cross section, we begin by writing the cross section of each individual run as
\begin{equation}
\sigma = \sigma_{+} + \sigma_{-} = \frac{wm_{1}}{N} -\frac{wm_{2}}{N},
\end{equation}
where $N$ denotes the total number of events in the LHE file, $m_{1}$ is the total number of positive weight events, $m_{2}$ the number of negative weights, and $w$ is absolute value of the weight. For the case where no cuts are imposed, we have that $m_{1} + m_{2} = N$, but this is not the case in general. Taking $\frac{\delta \sigma_{+}}{\sigma_{+}} = \frac{1}{\sqrt{m_1}}$ and similar for $\sigma_{-}$, we have that

\begin{equation}
\begin{split}
\delta  \sigma^2 = \delta\sigma_+^2 + \delta \sigma_-^2 = \frac{w^2(m_1+m_2)}{N^2} \\
\delta \sigma = \sigma \frac{\sqrt{m_1+m_2}}{m_1-m_2} \label{erroreq}
\end{split}
\end{equation}
which is the error for each LHE file. The second line of Eq. \eqref{erroreq} assumes that $m_1 \neq m_2$. Since the total cross section of a given bin is found by summing up the individual cross sections of the LHE files, we have that
\begin{equation}
\delta \sigma_{bin}^2 = \sum_i \delta \sigma_{i}^2
\end{equation}

\begin{table}[H]
\small
\caption{ The SM cross sections in [fb]  in the two dimensional bins $m_{t\bar t}, \theta^*$ for the process $\mu^+ \mu^- \rightarrow t \bar t \nu \bar \nu$ after all the preliminary cuts in Eq.~(\ref{eq:recoil}), (\ref{eq:thetacut}), (\ref{eq:etcut1}), (\ref{eq:etcut2}).}
\label{tab:xs10}
\hskip-1.75cm
\tabcolsep 10pt\begin{tabular}{|c|P{2.0cm}|P{2.0cm}|P{2.0cm}|P{2.0cm}|P{2.0cm}|P{2.0cm}|}
\hline
$m_{ t \bar t} [\text{TeV}] $ / $\theta^*$ [$^\circ$]  & \multirow{2}{*}{\textbf{[0,30]} }&  \multirow{2}{*}{\textbf{[30,60]}} &  \multirow{2}{*}{\textbf{[60,90]}}& \multirow{2}{*}{\textbf{[90,120]}}&  \multirow{2}{*}{\textbf{[120,150]}} &  \multirow{2}{*}{\textbf{[150,180]}}\\
\text{[fb]} & & & & & & \\
\hline
\textbf{[0, 1]} & 0.670 $\pm$ 0.00025 & 1.22 $\pm$ 0.00039 & 1.48 $\pm$ 0.00049 & 0.503 $\pm$ 0.00038 & 0.0933 $\pm$ 0.00022 & 0.0145 $\pm$ 0.00011 \\

\hline
\textbf{[1, 2]} & 0.234 $\pm$ $8.5 \times 10^{-5}$ & 0.233 $\pm$ 0.00012 & 0.142 $\pm$ 0.00012 & 0.0403 $\pm$ $8.3 \times 10^{-5}$ & 0.0122 $\pm$ $5.8 \times 10^{-5}$ & 0.00270 $\pm$ $3.5 \times 10^{-5}$ \\

\hline
\textbf{[2, 4]}  &$0.0449 \pm 2.3 \times 10^{-5}$  & $0.0322 \pm 2.6 \times 10^{-5}$ & $0.0141  \pm 2.3 \times 10^{-5}$ & $4.61 \times 10^{-3} \pm 1.6 \times 10^{-5}$ & $1.95 \times 10^{-3} \pm 1.3 \times 10^{-5}$ & $6.52 \times 10^{-4} \pm 9.0 \times 10^{-6}$ \\

\hline

\textbf{[4, 6]} & $3.08 \times 10^{-3} \pm 2.2 \times 10^{-6}$ & $1.76 \times 10^{-3} \pm 2.3 \times 10^{-6}$ & $6.49 \times 10^{-4} \pm 1.8 \times 10^{-6}$ & $2.78 \times 10^{-4} \pm 1.4 \times 10^{-6}$ & $1.40 \times 10^{-4} \pm 1.1 \times 10^{-6}$ & $8.01 \times 10^{-5} \pm 8.2 \times 10^{-7}$ \\
\hline

\textbf{[6, 8]}  & $2.46 \times 10^{-4} \pm 1.9 \times 10^{-7}$ & $1.23 \times 10^{-4} \pm 1.9 \times 10^{-7}$ & $5.56 \times 10^{-5} \pm 1.5 \times 10^{-7}$ & $3.16 \times 10^{-5} \pm 1.1 \times 10^{-7}$ & $2.43 \times 10^{-5} \pm 8.8\times 10^{-8}$ & $1.73 \times 10^{-5} \pm 6.9 \times 10^{-8}$ \\

\hline
\textbf{[8, 10]}  & $8.33 \times 10^{-6} \pm 5.2 \times 10^{-9}$ & $6.71 \times 10^{-6} \pm 5.2 \times 10^{-9}$ & $6.47 \times 10^{-6} \pm 3.9 \times 10^{-9}$ & $6.21 \times 10^{-6} \pm 3.0 \times 10^{-9}$ & $7.13 \times 10^{-6} \pm 2.4 \times 10^{-9}$ & $9.54 \times 10^{-6} \pm 1.9 \times 10^{-9}$ \\

\hline
\end{tabular}
\end{table}

\begin{table}[H]
\small
\caption{ The SM cross sections  in the two dimensional bins $m_{t\bar t}, \theta^*$ for the process $\mu^+ \mu^- \rightarrow t \bar t \nu \bar \nu$ at 30 TeV muon collider after all the preliminary cuts in Eq.~(\ref{eq:recoil}), (\ref{eq:thetacut}), (\ref{eq:etcut1}), (\ref{eq:etcut2}).}
\label{tab:xs30}
\hskip-0.75cm
\tabcolsep 6pt\begin{tabular}{|c|P{2.0cm}|P{2.0cm}|P{2.0cm}|P{2.0cm}|P{2.0cm}|P{2.0cm}|}
\hline
$m_{ t \bar t} [\text{TeV}] $ / $\theta^*$ [$^\circ$]  & \multirow{2}{*}{\textbf{[0,30]} }&  \multirow{2}{*}{\textbf{[30,60]}} &  \multirow{2}{*}{\textbf{[60,90]}}& \multirow{2}{*}{\textbf{[90,120]}}&  \multirow{2}{*}{\textbf{[120,150]}} &  \multirow{2}{*}{\textbf{[150,180]}}\\
\text{[fb]} & & & & & & \\
\hline
\textbf{[0, 1]} & 0.641 $\pm$ 0.00033 & 1.28 $\pm$ 0.00055 & 1.95 $\pm$ 0.00080 & 0.61 $\pm$ 0.00060 & 0.085 $\pm$ 0.00031 & 0.0119 $\pm$ 0.00014 \\

\hline
\textbf{[1, 5]} & 0.368 $\pm$ 0.00021 & 0.376 $\pm$ 0.00026 & 0.314 $\pm$ 0.00031 & 0.0821 $\pm$ 0.00022 & 0.0174 $\pm$ 0.00013 & 0.00375 $\pm$ 0.000083 \\

\hline
\textbf{[5, 10]}  &$8.33 \times 10^{-3}\pm 1.2 \times 10^{-5}$  & $6.40 \times 10^{-3} \pm 1.4 \times 10^{-5}$ & $3.03 \times 10^{-3} \pm 1.3 \times 10^{-5}$ & $9.68 \times 10^{-4} \pm 9.0 \times 10^{-6}$ & $3.90 \times 10^{-4} \pm 6.6 \times 10^{-6}$ & $1.27 \times 10^{-4} \pm 4.8 \times 10^{-6}$ \\

\hline

\textbf{[10, 15]} & $7.52 \times 10^{-4} \pm 1.5 \times 10^{-6}$ & $4.55 \times 10^{-4} \pm 1.7 \times 10^{-6}$ & $1.73 \times 10^{-4} \pm 1.3 \times 10^{-6}$ & $6.71 \times 10^{-5} \pm 9.8 \times 10^{-7}$ & $3.22 \times 10^{-5} \pm 7.7 \times 10^{-7}$ & $1.41 \times 10^{-5} \pm 5.8 \times 10^{-7}$ \\
\hline

\textbf{[15, 20]}  & $9.77 \times 10^{-5} \pm 2.2 \times 10^{-7}$ & $5.02 \times 10^{-5} \pm 2.3 \times 10^{-7}$ & $1.97 \times 10^{-5} \pm 1.8 \times 10^{-7}$ & $9.15 \times 10^{-6} \pm 1.3 \times 10^{-7}$ & $5.65 \times 10^{-6} \pm 1.1\times 10^{-7}$ & $3.17 \times 10^{-6} \pm 8.3 \times 10^{-8}$ \\

\hline
\textbf{[20, 25]}  & $1.02 \times 10^{-5} \pm 2.4 \times 10^{-8}$ & $5.29 \times 10^{-6} \pm 2.4 \times 10^{-8}$ & $2.80 \times 10^{-6} \pm 1.8 \times 10^{-8}$ & $1.83 \times 10^{-6} \pm 1.4 \times 10^{-8}$ & $1.46 \times 10^{-6} \pm 1.1 \times 10^{-8}$ & $1.12 \times 10^{-6} \pm 8.5 \times 10^{-9}$ \\

\hline
\textbf{[25, 30]} & $4.47 \times 10^{-7} \pm 7.2 \times 10^{-10}$ & $4.81 \times 10^{-7} \pm 7.2 \times 10^{-10}$ & $5.44 \times 10^{-7} \pm 5.4 \times 10^{-10}$ & $5.93 \times 10^{-7} \pm 4.1 \times 10^{-10}$ & $7.04 \times 10^{-7} \pm 3.3 \times 10^{-10}$ & $1.02 \times 10^{-6} \pm 2.6 \times 10^{-10}$ \\

\hline
\end{tabular}
\end{table}

\begin{table}[H]
\caption{ $R$-values  in the two dimensional bins $m_{t\bar t}, \theta^*$ for the process $\mu^+ \mu^- \rightarrow t \bar t \nu \bar \nu$ after all the preliminary cuts in Eq.~(\ref{eq:recoil}), (\ref{eq:thetacut}), (\ref{eq:etcut1}), (\ref{eq:etcut2}).}
\label{tab:rval10}
\hskip-1.85cm
\tabcolsep 10pt\begin{tabular}{|c|P{2.0cm}|P{2.0cm}|P{2.0cm}|P{2.0cm}|P{2.0cm}|P{2.0cm}|}
\hline
$m_{ t \bar t}[\text{TeV}] $  / $\theta^*$[$^\circ$]  & \textbf{[0,30]} &  \textbf{[30,60]} &  \textbf{[60,90]}&  \textbf{[90,120]}&  \textbf{[120,150]} &  \textbf{[150,180]}\\
\hline
\textbf{[0, 1]} & & &  &  && \\
$R_1$& 0.209 $\pm$ 0.070 & 0.143 $\pm$ 0.052 & -0.0484 $\pm$ 0.047 & -0.647 $\pm$ 0.081 & -1.54 $\pm$ 0.19 & -2.60 $\pm$ 0.48 \\
$R_2$& 0.279 $\pm$ 0.67 & 0.563 $\pm$ 0.51 & 1.16 $\pm$ 0.48 & 3.40 $\pm$ 0.82 & 7.20 $\pm$ 1.7 & 12.8 $\pm$ 4.1 \\

\hline
\textbf{[1, 2]} & & &  &  && \\
$R_1$ & -0.0314 $\pm$ 0.055 & -0.0778 $\pm$ 0.055 & -0.266 $\pm$ 0.071 & -0.827 $\pm$ 0.13 & -1.79 $\pm$ 0.24 & -3.81 $\pm$ 0.51 \\
$R_2$& 1.29 $\pm$ 0.57 & 3.06 $\pm$ 0.54 & 8.43 $\pm$ 0.70 & 29.9 $\pm$ 1.4 & 58.3 $\pm$ 2.3 & 110 $\pm$ 4.5 \\

\hline
\textbf{[2, 4]} & & & & & & \\
$R_1$& -0.0124 $\pm$ 0.052 & -0.116 $\pm$ 0.061 & -0.313 $\pm$ 0.092 & -0.790 $\pm$ 0.16 & -1.33 $\pm$ 0.25 & -2.68 $\pm$ 0.43 \\
$R_2$ & 5.28 $\pm$ 0.54 & 12.8 $\pm$ 0.57 & 36.2 $\pm$ 0.89 & 110 $\pm$ 1.7 & 212 $\pm$ 2.7 & 362 $\pm$ 6.2 \\

\hline
\textbf{[4, 6]} & & & & & & \\
$R_1$ & -0.0109 $\pm$ 0.050 & -0.127 $\pm$ 0.066 & -0.331 $\pm$ 0.11 & -0.481 $\pm$ 0.17 & -0.842 $\pm$ 0.23 & -1.25 $\pm$ 0.31 \\
$R_2$& 17.1 $\pm$ 0.53 & 44.7 $\pm$ 0.60 & 128 $\pm$ 1.1 & 297 $\pm$ 2.2 & 556 $\pm$ 4.9 & 649 $\pm$ 7.1 \\

\hline

\textbf{[6, 8]} & & &  &  && \\
$R_1$& -0.0171 $\pm$ 0.049 & -0.0997 $\pm$ 0.069 & -0.205 $\pm$ 0.10 & 0.211 $\pm$ 0.14 & -0.252 $\pm$ 0.16 & -0.218 $\pm$ 0.18 \\
$R_2$ & 33.7 $\pm$ 0.52 & 95.5 $\pm$ 0.65 & 207 $\pm$ 1.1 & 365 $\pm$ 1.9 & 478 $\pm$ 2.4 & 479 $\pm$ 2.5 \\

\hline
\textbf{[8, 10]} & & & & & & \\
$R_1$ & 0.0116 $\pm$ 0.073 & -0.0316 $\pm$ 0.082 & -0.0188 $\pm$ 0.083 & -0.0103 $\pm$ 0.085 & 0.00346 $\pm$ 0.079 & 0.00128 $\pm$ 0.068 \\
$R_2$ & 47.9 $\pm$ 0.79 & 81.3 $\pm$ 0.77 & 82.1 $\pm$ 0.80 & 85.2 $\pm$ 0.88 & 76.7 $\pm$ 0.86 & 41.7 $\pm$ 0.62 \\

\hline
\end{tabular}
\end{table}

\begin{table}[H]
\caption{ $R$-values  in the two dimensional bins $m_{t\bar t}, \theta^*$ for the process $\mu^+ \mu^- \rightarrow t \bar t \nu \bar \nu$ after all the preliminary cuts in Eq.~(\ref{eq:recoil}), (\ref{eq:thetacut}), (\ref{eq:etcut1}), (\ref{eq:etcut2}).}
\label{tab:rval30}
\hskip-0.9cm
\tabcolsep 6pt\begin{tabular}{|c|P{2.0cm}|P{2.0cm}|P{2.0cm}|P{2.0cm}|P{2.0cm}|P{2.0cm}|}
\hline
$m_{ t \bar t}[\text{TeV}] $  / $\theta^*$[$^\circ$]  & \textbf{[0,30]} &  \textbf{[30,60]} &  \textbf{[60,90]}&  \textbf{[90,120]}&  \textbf{[120,150]} &  \textbf{[150,180]}\\
\hline
\textbf{[0, 1]} & & &  &  && \\
$R_1$& 0.197 $\pm$ 0.0056 & 0.137 $\pm$ 0.0044 & -0.0365 $\pm$ 0.0041 & -0.694 $\pm$ 0.012 & -1.66 $\pm$ 0.057 & -2.88 $\pm$ 0.24 \\
$R_2$& 0.286 $\pm$ 0.0078 & 0.508 $\pm$ 0.0072 & 1.08 $\pm$ 0.0091 & 3.55 $\pm$ 0.042 & 9.70 $\pm$ 0.32 & 22.3 $\pm$ 2.1 \\

\hline
\textbf{[1, 5]} & & &  &  && \\
$R_1$ & -0.00399 $\pm$ 0.0057 & -0.0849 $\pm$ 0.0071 & -0.287 $\pm$ 0.0099 & -0.956 $\pm$ 0.030 & -2.05 $\pm$ 0.10 & -4.76 $\pm$ 0.39 \\
$R_2$& 2.66 $\pm$ 0.028 & 5.50 $\pm$ 0.049 & 12.8 $\pm$ 0.11 & 48.0 $\pm$ 0.71 & 145 $\pm$ 5.2 & 368 $\pm$ 30 \\

\hline
\textbf{[5, 10]} & & & & & & \\
$R_1$& -0.0171 $\pm$ 0.015 & -0.101 $\pm$ 0.022 & -0.236 $\pm$ 0.041 & -0.824 $\pm$ 0.095 & -0.882 $\pm$ 0.17 & -2.59 $\pm$ 0.41 \\
$R_2$ & 32.4 $\pm$ 0.30 & 92.4 $\pm$ 0.94 & 242 $\pm$ 3.3 & 654 $\pm$ 14 & 1459 $\pm$ 54 & 3465 $\pm$ 259 \\

\hline
\textbf{[10, 15]} & & & & & & \\
$R_1$ & -0.00494 $\pm$ 0.020 & -0.131 $\pm$ 0.037 & -0.346 $\pm$ 0.077 & -0.732 $\pm$ 0.15 & -1.09 $\pm$ 0.24 & -1.35 $\pm$ 0.42 \\
$R_2$& 99.5 $\pm$ 0.84 & 335 $\pm$ 3.6 & 890 $\pm$ 14 & 1946 $\pm$ 45 & 4012 $\pm$ 138 & 7773 $\pm$ 460 \\

\hline

\textbf{[15, 20]} & & &  &  && \\
$R_1$& -0.0605 $\pm$ 0.023 & -0.129 $\pm$ 0.046 & -0.351 $\pm$ 0.092 & 0.406 $\pm$ 0.15 & -0.914 $\pm$ 0.19 & -0.639 $\pm$ 0.26 \\
$R_2$ & 195 $\pm$ 1.5 & 738 $\pm$ 7.8 & 1770 $\pm$ 28 & 3159 $\pm$ 65 & 5080 $\pm$ 133 & 9193 $\pm$ 386 \\

\hline
\textbf{[20, 25]} & & & & & & \\
$R_1$ & -0.0195 $\pm$ 0.023 & -0.0786 $\pm$ 0.045 & -0.0399 $\pm$ 0.064 & -0.0314 $\pm$ 0.074 & -0.201 $\pm$ 0.075 & -0.268 $\pm$ 0.076 \\
$R_2$ & 321 $\pm$ 2.5 & 1166 $\pm$ 13 & 1942 $\pm$ 27 & 2519 $\pm$ 40 & 3201 $\pm$ 56 & 4070 $\pm$ 95 \\

\hline
\textbf{[25, 30]} & & & & & & \\
$R_1$ & 0.0111 $\pm$ 0.016 & -0.00137 $\pm$ 0.015 & -0.0253 $\pm$ 0.010 & -0.00295$\pm$ 0.0069 & 0.00193 $\pm$ 0.0047 & 0.00218 $\pm$ 0.0026 \\
$R_2$& 339 $\pm$ 4.3 & 568 $\pm$ 7.7 & 488 $\pm$ 6.3 & 373 $\pm$ 4.3 & 303 $\pm$ 3.1 & 214 $\pm$ 2.2 \\
\hline
\end{tabular}
\end{table}

\section{Helicity Amplitudes for $W^+ W^- \rightarrow t \bar{t}$}
\label{app:hcamp}
In this appendix, we present the full helicity amplitudes for the subprocess $W^+ W^- \rightarrow t \bar{t}$:
\beq
\begin{split}
 \mathcal{M} (W^+(p_1) W^-(p_2)  \rightarrow t(p_3)\bar{t}(p_4)  )& = \mathcal{M}^{\gamma} + \mathcal{M}^{Z}+ \mathcal{M}^h + \mathcal{M}^{t}\\
\end{split}
\eeq
where $\mM^{\gamma, Z, h}$ denotes the $s-$channel contribution with $\gamma, Z, h$  particles as internal lines and $\mM^t$ corresponds to the $t-$channel contribution. Since the initial particles have the same masses as well as the final particles, the energies of the top quarks are equal to that of the $W$ bosons in the partonic center-of-mass frame:
\beq
\hat E_t = \hat E_W = \frac{\sqrt{\hat s}}{2}
\eeq
The other Mandelstam variables $\hat t, \hat u$ can be written as functions of $\hat s$:
\beq
 \hat t= \frac{\hat s}{4}\left(- \beta_t^2 - \beta_W^2 +  2 \beta_t \beta_W \cos\theta\right), \qquad \hat u = \frac{\hat s}{4}\left(- \beta_t^2 - \beta_W^2 -  2 \beta_t \beta_W \cos\theta\right),
\eeq
where the velocities of the $W$-bosons and the top quarks are given by:
\beq
\beta_{W,t} = \sqrt{1- \frac{4m_{W,t}^2}{\hat s}}
\eeq
Here the scattering angle $\theta$ in the partonic center-of-mass frame is the polar angle between the out-going top quark and the incoming $W^{+}$ gauge boson. The $z$-axis in chosen the direction of the $W^+$ spatial momentum. The azimuthal angles of the top quark and the anti-top quark are chosen as:
\beq
\varphi_t = 0, \qquad \varphi_{\bar t} = \pi
\eeq
which will fix the possible $i$ factors in the polarization functions of the anti-top quarks.
We will present the helicity amplitudes in terms of the Wigner $d$ functions~\cite{Hagiwara:1986vm}:
\beq
\mM_{h_1 h_2; h_3 h_4} = \widetilde{\mM}_{h_1 h_2; h_3 h_4} (\theta) (h_3- h_4+\delta_{h_3  h_4}) (-1)^{h_2} d_{\Delta h_{12}, \Delta h_{34}}^{J_0}(\theta)
\eeq
with
\beq
\Delta h_{12}= h_{1}- h_{2}, \qquad \Delta h_{34} = h_3 - h_4, \qquad J_0 = \text{max}(| \Delta h_{12}|, | \Delta h_{34}|)
\eeq
and to make results more compact, we have also extracted some sign factors for convenience. The relevant $d$ functions are listed as follows~\cite{Workman:2022ynf}:
\beq
\begin{split}
d_{1,1}^1 &= d_{-1,-1}^1  = \frac12  (1 + \cos\theta),\qquad d_{1,-1}^1 = d_{-1,1}^1  = \frac12  (1 - \cos\theta), \\ 
d_{1,0}^1 &= -d_{-1,0}^1  =- \frac{\sin\theta}{\sqrt{2}}\\
d_{1,2}^2 &=- d_{-1,-2}^2  = \frac12 \sin\theta (1 + \cos\theta),\qquad d_{1,-2}^2 =- d_{-1,2}^2  =- \frac12 \sin\theta (1 - \cos\theta)\\
\end{split}
\eeq
which satisfy the following identities:
\beq
d^j_{m^\prime, m} = (-1)^{m-m^\prime} d^j_{m,m^\prime} = d^j_{-m,-m^\prime}
\eeq
The top Yukawa coupling modification is parametrized as:
\beq
\mL_{ht\bar t} = - \frac{m_t}{v} (1 + \delta_{tth}) h t \bar t
\eeq
 For future studies, we have also included the $CP$-even anomalous triple gauge boson couplings (aTGC), which are parametrized as follows~\cite{Hagiwara:1986vm}:
\begin{equation}
\label{eq:cubic}
\begin{split}
\mathcal{L}_{WWV}/g^{\rm SM}_{WWV}  = 
&  i g_1^V\, ( W^+_{\mu\nu} W^{-\mu }V^{\nu}  
                                 -  W^-_{\mu\nu} W^{+\mu }  V^{ \nu } )+  i \kappa_V W^+_\mu  W^-_\nu  V^{\mu\nu} + i \frac{\lambda_V}{m_W^2} W^+_{\lambda \mu} W^{-\mu}_{ \  \ \  \  \nu} V^{\nu \lambda}\\
\end{split}
\end{equation}
where  $W^{\pm}_{\mu\nu} = \partial_\mu W_\nu^\pm - \partial_\nu W_\mu^\pm $ and V = $\gamma$, Z. The SM values of the TGCs read:
\beq
g^{\rm SM}_{WW\gamma} = e, \qquad g^{\rm SM}_{WWZ} = g \cos\theta_W.
\eeq 
where $\theta_W$ is the weak mixing angle.
The unbroken electromagnetism  fixes $g_1^\gamma$  to be 1.
So we are left with 5 anomalous  TGC couplings: $\delta g_1^Z,  \delta \kappa_Z, \delta \kappa_\gamma, \lambda_Z, \lambda_\gamma$ defined as $\delta g_1^Z = g_1^Z - 1 , \delta \kappa_V = \kappa_V - 1$. At dimension-six SMEFT, they are further related by the following identities~\cite{Giudice:2007fh}:
\beq
\delta \kappa_Z = \delta g_1^Z - \tan^2{\theta_W} \delta \kappa_\gamma, \qquad \lambda_Z = \lambda_\gamma.
\eeq
but here we will take them as independent couplings. We also take into account the contributions from the possible modifications of the top electroweak couplings and the Higgs gauge boson coupling:
\beq
\delta_{Wtb} = \frac{g_{Wtb}}{g_{Wtb}^{\rm SM}}-1, \qquad \delta_{Zt_L} = \frac{g_{Zt_L}}{g_{Zt_L}^{\rm SM}}-1, \qquad \delta_{Zt_R} = \frac{g_{Zt_R}}{g_{Zt_R}^{\rm SM}}-1, \qquad \delta_{hWW} = \frac{g_{hWW}}{g^{\rm SM}_{hWW}} - 1
\eeq
with their SM values as follows:
\beq
g_{Wt}^{\rm SM} = \frac{g}{\sqrt{2}}, \quad  g_{Zt_L}^{\rm SM} = \frac{g}{\cos\theta_W} \left(\frac12 - \frac 23 \sin^2\theta_W\right), \quad    g_{Zt_R}^{\rm SM} = -  \frac 23  \frac{g \sin^2\theta_W}{\cos\theta_W} , \quad  g^{\rm SM}_{hWW} = \frac{2m_W^2}{v}
\eeq 
Now, we turn to the  formulae for the helicity amplitudes. In order to list them compactly in tables, we further take some pre-factors out of $\widetilde \mM$:
\beq
\begin{split}
\widetilde{\mM}^\gamma &= i \frac{2 \sqrt{2} g^2 s_W^2 \beta_W}{3} A^{\gamma}_{h_1 h_2; h_3 h_4}\\
\widetilde{\mM}^Z &= i \sqrt{2} g^2  \beta_W\left(\frac{1- \Delta h_{34}\,\beta_t}{4}(1+\delta_{Zt_L})  - \frac{2}{3}s_W^2\left(1 + \frac{1- \Delta h_{34}\,\beta_t}{2}\delta_{Zt_L} + \frac{1+\Delta h_{34}\,\beta_t}{2}\delta_{Zt_R} \right)\right) \times \\
&\quad \frac{\hat s}{\hat s-m_Z^2}A^{Z}_{h_1 h_2; h_3 h_4}\\
\widetilde{\mM}^h &= i \frac{g^2}{2\sqrt{2}} (1+\delta_{tth}) (1 +\delta _{hWW})\beta_t \frac{\hat s}{\hat s-m_h^2}A^{h}_{h_1 h_2; h_3 h_4}\\
\widetilde{\mM}^t &= -i \frac{ g^2(1-\Delta h_{34} \, \beta_t)}{2\sqrt{2} \beta_W} (1 + \delta_{Wtb})^2\left(B_{h_1 h_2; h_3 h_4} - \frac{1}{\beta_t^2 + \beta_W^2 - 2 \beta_t \beta_W \cos\theta }C_{h_1 h_2; h_3 h_4}\right)\\
\end{split}
\eeq
where we have abbreviated  $\sin\theta_W$ as $s_W$.
Note that the kinematical function in front of $C_{h_1,h_2;h_3,h_4}$ is simply $\frac{\hat s}{4\hat t}$ and we have omitted the small bottom quark mass.  The results for  the helicity configurations $(\mp \frac12, \pm \frac 12)$ of final top and anti-top quarks are presented in Table~\ref{tab:hc1} and for other helicity configurations $(\mp \frac12, \mp \frac 12)$, they are shown in Table~\ref{tab:hc2}. 


\begin{table}[H]
\small
\begin{center}
\caption{Helicity amplitude factors for $W_{h_1}^+ W_{h_2}^- \rightarrow t_{h_3} \bar t_{h_4} $ for $\Delta h_{34} = \mp 1$. Here $V = \gamma, Z$ and note that $\delta g_1^\gamma = 0$.} 
\label{tab:hc1}
\begin{tabular}{|c|c|c|c|c|c|c|c|c|}
\hline
$(h_3 h_4)$ & $(h_1 h_2)$& $A^{V}_{h_1 h_2;  h_3 h_4}$     & $A^{h}_{h_1 h_2;  h_3 h_4}$ &  $B_{h_1 h_2;  h_3 h_4}$ & $C_{h_1 h_2;  h_3 h_4}$  \\
  \hline
\multirow{5}{*}{$(-\frac12\ \frac12)$}        & (+1 -1),(-1 +1)   &0 & 0 &0&$-2 \sqrt{2} \beta_t\beta_W$\\
     &  (+1 +1),(-1 -1)  & $ 1+ \delta g_1^V+\frac{s}{2 m_W^2}   \lambda_V$  &0  & $1$  & $\beta_t^2-\beta_W^2$\\
    & (+1 0),(0 -1)   & $ \frac{\sqrt{s}}{m_W}\left(1+ \frac{\delta g_1^V + \delta\kappa_V+\lambda_V}{2}\right)$  &0 & $\frac{\sqrt{s}}{m_W}$&$\frac{\sqrt{s}(\beta_t+\beta_W)(\beta_t -\beta_W^2)}{m_W}$\\
     & (-1 0),(0 +1)   & $-\frac{\sqrt{s}}{m_W}\left(1+ \frac{\delta g_1^V +\delta\kappa_V+\lambda_V}{2}\right)$&0& $-\frac{\sqrt{s}}{m_W}$& $-\frac{\sqrt{s}(\beta_t-\beta_W)(\beta_t -\beta_W^2)}{m_W}$  \\
      & (0 0)& $ -1- \delta g_1^V -\frac{s}{2 m_W^2} (1 + \delta \kappa_V)$ &0 & $-\frac{s}{2m_W^2}$& $-\frac{s\left(\beta_t -\beta_W^2\right)^2}{2m_W^2}$ \\
  \hline
\multirow{5}{*}{$(\frac12\ -\frac12)$}        & (+1 -1),(-1 +1)   &0 & 0 &0&$-2 \sqrt{2} \beta_t\beta_W$\\
     &  (+1 +1),(-1 -1)  & $ 1+\delta g_1^V+\frac{s}{2 m_W^2}   \lambda_V$  &0  & $1$  & $\beta_t^2-\beta_W^2$\\
    & (+1 0),(0 -1)   & $ \frac{\sqrt{s}}{m_W}\left(1+ \frac{\delta g_1^V+\delta\kappa_V+\lambda_V}{2}\right)$  &0 & $\frac{\sqrt{s}}{m_W}$&$\frac{\sqrt{s}(\beta_t-\beta_W)(\beta_t +\beta_W^2)}{m_W}$\\
     & (-1 0),(0 +1)   & $-\frac{\sqrt{s}}{m_W}\left(1+ \frac{\delta g_1^V+\delta\kappa_V+\lambda_V}{2}\right)$&0& $-\frac{\sqrt{s}}{m_W}$& $-\frac{\sqrt{s}(\beta_t+\beta_W)(\beta_t +\beta_W^2)}{m_W}$  \\
      & (0 0)& $ -1-\delta g_1^V-\frac{s}{2 m_W^2} (1 + \delta \kappa_V)$ &0 & $-\frac{s}{2m_W^2}$& $-\frac{s\left(\beta_t +\beta_W^2\right)^2}{2m_W^2}$ \\
\hline
\end{tabular}
\end{center}
\end{table}

\begin{table}[H]
\small
\begin{center}
\caption{ Helicity amplitude factors for $W_{h_1}^+ W_{h_2}^- \rightarrow t_{h_3} \bar t_{h_4} $ for $\Delta h_{34} = 0$. Here $V = \gamma, Z$ and note that $\delta g_1^\gamma = 0$.}
\label{tab:hc2}
\begin{tabular}{|c|c|c|c|c|c|c|c|c|}
\hline
$(h_3 h_4)$ & $(h_1 h_2)$& $A^{V}_{h_1 h_2;  h_3 h_4}$     & $A^{h}_{h_1 h_2;  h_3 h_4}$  \\
\hline
\multirow{5}{*}{$(-\frac12\ -\frac12)$}        & (+1 -1),(-1 +1)   &0 &0   \\
     &  (+1 +1),(-1 -1)  & $ \frac{\sqrt{2}m_t}{\sqrt{s}}\left(1+\delta g_1^V+\frac{s}{2 m_W^2}   \lambda_V\right) \cos\theta$ & $-\frac{\sqrt{2}m_t}{\sqrt{s}}$   \\
    & (+1 0),(0 -1)   & $ \frac{\sqrt{2}m_t}{m_W}\left(1+ \frac{\delta g_1^V+\delta\kappa_V+\lambda_V}{2}\right)$  &0  \\
     & (-1 0),(0 +1)   & $- \frac{\sqrt{2}m_t}{m_W}\left(1+ \frac{\delta g_1^V+\delta\kappa_V+\lambda_V}{2}\right)$ & 0 \\
      & (0 0)&  $ -\frac{\sqrt{2}m_t}{\sqrt{s}}\left(1+\delta g_1^V+\frac{s  (1+ \delta\kappa_V)}{2 m_W^2} \right) \cos\theta$ & $-\frac{\sqrt{2}m_t \sqrt{s}(1+\beta_W^2)}{4 m_W^2}$  \\
      \hline
\multirow{5}{*}{$(\frac12\ \frac12)$}        & (+1 -1),(-1 +1)   &0 &0  \\
     &  (+1 +1),(-1 -1)  & $- \frac{\sqrt{2}m_t}{\sqrt{s}}\left(1+\delta g_1^V+\frac{s}{2 m_W^2}   \lambda_V\right) \cos\theta$ & $\frac{\sqrt{2}m_t}{\sqrt{s}}$   \\
    & (+1 0),(0 -1)   & $- \frac{\sqrt{2}m_t}{m_W}\left(1+ \frac{\delta g_1^V+\delta\kappa_V+\lambda_V}{2}\right)$  &0   \\
     & (-1 0),(0 +1)   & $ \frac{\sqrt{2}m_t}{m_W}\left(1+ \frac{\delta g_1^V+\delta\kappa_V+\lambda_V}{2}\right)$ & 0  \\
      & (0 0)&  $ \frac{\sqrt{2}m_t}{\sqrt{s}}\left(1+\delta g_1^V+\frac{s  (1+ \delta\kappa_V)}{2 m_W^2} \right) \cos\theta$ & $\frac{\sqrt{2}m_t \sqrt{s}(1+\beta_W^2)}{4 m_W^2}$ \\
      \hline
\end{tabular}
\begin{tabular}{|c|c|c|c|c|c|}
      \hline
$(h_3 h_4)$ & $(h_1h_2)$ &  $B_{h_1 h_2;  h_3 h_4}$& $C_{h_1 h_2;  h_3 h_4}$  \\
\hline
\multirow{5}{*}{$(-\frac12\ -\frac12)$}        & (+1 -1),(-1 +1)   &0&$-\frac{8 m_t \beta_t\beta_W}{\sqrt{3}\sqrt{s}}$ \\
     &  (+1 +1),(-1 -1) &$\frac{m_t(\beta_t^2-\beta_W^2\mp 2 \beta_t \beta_W+ 2 \beta_t \beta_W\cos\theta)}{\sqrt{2}\sqrt{s}\beta_t\beta_W}$ & $\frac{m_t(\beta_t\mp\beta_W)^2(\beta_t^2-\beta_W^2)}{\sqrt{2}\sqrt{s}\beta_t\beta_W}$  \\
    & (+1 0),(0 -1)  &$\frac{\sqrt{2}m_t}{m_W}$  &$\frac{\sqrt{2}m_t(\beta_t\mp\beta_W)(\beta_t\pm\beta_W^2)}{m_W}$ \\
     & (-1 0),(0 +1)    &$-\frac{\sqrt{2}m_t}{m_W}$  &$-\frac{\sqrt{2}m_t(\beta_t\pm\beta_W)(\beta_t\pm\beta_W^2)}{m_W}$  \\
      & (0 0)& $-\frac{\sqrt{s}m_t (\beta_t^2 + \beta_W^4+2 \beta_t \beta_W\cos\theta)}{2\sqrt{2}m_W^2\beta_t\beta_W}$&$-\frac{\sqrt{s}m_t(\beta_t^4+\beta_W^6-\beta_t^2(\beta_W^2 +\beta_W^4))}{2\sqrt{2}m_W^2\beta_t\beta_W}$\\
      \hline
\multirow{5}{*}{$(\frac12\ \frac12)$}        & (+1 -1),(-1 +1) &0   &$\frac{8 m_t \beta_t\beta_W}{\sqrt{3}\sqrt{s}}$ \\
     &  (+1 +1),(-1 -1)&$-\frac{m_t(\beta_t^2-\beta_W^2\pm 2 \beta_t \beta_W+ 2 \beta_t \beta_W\cos\theta)}{\sqrt{2}\sqrt{s}\beta_t\beta_W}$ & $-\frac{m_t(\beta_t\pm\beta_W)^2(\beta_t^2-\beta_W^2)}{\sqrt{2}\sqrt{s}\beta_t\beta_W}$  \\
    & (+1 0),(0 -1)  &$-\frac{\sqrt{2}m_t}{m_W}$ &$-\frac{\sqrt{2}m_t(\beta_t\pm\beta_W)(\beta_t\mp\beta_W^2)}{m_W}$ \\
     & (-1 0),(0 +1)  &$\frac{\sqrt{2}m_t}{m_W}$  &$\frac{\sqrt{2}m_t(\beta_t\mp\beta_W)(\beta_t\mp\beta_W^2)}{m_W}$  \\
      & (0 0) & $\frac{\sqrt{s}m_t (\beta_t^2 + \beta_W^4+2 \beta_t \beta_W\cos\theta)}{2\sqrt{2}m_W^2\beta_t\beta_W}$&$\frac{\sqrt{s}m_t(\beta_t^4+\beta_W^6-\beta_t^2(\beta_W^2 +\beta_W^4))}{2\sqrt{2}m_W^2\beta_t\beta_W}$\\
\hline
\end{tabular}
\end{center}
\end{table}

\section{Statistics}
\label{app:stat}

In order to constrain the top Yukawa coupling as shown in Fig.~\ref{rvalsplot}, we follow the frequentist statistics procedure outlined in \cite{ParticleDataGroup:2020ssz}. We first construct the likelihood function $L(\delta_{tth})$:
\beq
L(\delta_{tth}) = P(n|\delta_{tth})
\eeq 
where $n$ is the observed number of events and $ P(n|\delta_{tth})$ is the probability under the hypothesis of $\delta_{tth}$. Here we have used the Poisson distribution:
\begin{equation}
\begin{split}
P(n|\delta_{tth}) = \frac{( s(\delta_{tth}) + b)^n}{n!} e^{-( s(\delta_{tth}) + b)}.\\
\end{split}
\end{equation}
where $s$ is the number of signal events, which is a function of $\delta_{tth}$ and $b$ is the  number of SM background events. For multi-bin analysis, as is the case in this paper, the total probability function is given by the product of the probability function in each bin, i.e.:
\beq
P(\vec n|\delta_{tth}) =\prod_i \frac{( s_i(\delta_{tth}) + b_i)^{n_i}}{n_i!} e^{-( s_i(\delta_{tth}) + b_i)}.\\
\eeq
The $ \chi^2$ function is defined as:
\beq
\chi^2 = - 2 \ln L
\eeq
and we will  use the method of maximum likelihood to estimate the confidence interval. The $\Delta \chi^2$ as plotted in Fig.~\ref{rvalsplot} is defined as:
\beq
\Delta \chi^2 = \chi^2 - \chi^2_{\rm min} = 2 \ln L_{\rm max} - 2 \ln L
\eeq
where  $L_{\rm max} $ is the maximal value of the likelihood function with given date $\vec n$.
The expected sensitivity is obtained by setting the observed number of events to the SM background values $\vec n = \vec b$. The confidence interval at  $m$-standard deviation is obtained by solving the following equation:
\beq
\Delta \chi^2 = m^2
\eeq


\bibliographystyle{utphys}
\bibliography{references2}

\providecommand{\href}[2]{#2}\begingroup\raggedright\begin{thebibliography}{10}

\bibitem{ALEPH:2005ab}
{\bfseries ALEPH, DELPHI, L3, OPAL, SLD, LEP Electroweak Working Group, SLD
  Electroweak Group, SLD Heavy Flavour Group} Collaboration, S.~Schael {\em
  et~al.}, ``{Precision electroweak measurements on the $Z$ resonance},''
  \href{http://dx.doi.org/10.1016/j.physrep.2005.12.006}{{\em Phys. Rept.}
  {\bfseries 427} (2006) 257--454},
  \href{http://arxiv.org/abs/hep-ex/0509008}{{\ttfamily arXiv:hep-ex/0509008}}.

\bibitem{LHCHiggsCrossSectionWorkingGroup:2011wcg}
{\bfseries LHC Higgs Cross Section Working Group} Collaboration, S.~Dittmaier
  {\em et~al.}, ``{Handbook of LHC Higgs Cross Sections: 1. Inclusive
  Observables},'' \href{http://arxiv.org/abs/1101.0593}{{\ttfamily
  arXiv:1101.0593 [hep-ph]}}.

\bibitem{Cornwall:1974km}
J.~M. Cornwall, D.~N. Levin, and G.~Tiktopoulos, ``{Derivation of Gauge
  Invariance from High-Energy Unitarity Bounds on the s Matrix},''
  \href{http://dx.doi.org/10.1103/PhysRevD.10.1145}{{\em Phys. Rev. D}
  {\bfseries 10} (1974) 1145}. [Erratum: Phys.Rev.D 11, 972 (1975)].

\bibitem{Liu:2022alx}
D.~Liu and Z.~Yin, ``{Gauge invariance from on-shell massive amplitudes and
  tree-level unitarity},''
  \href{http://dx.doi.org/10.1103/PhysRevD.106.076003}{{\em Phys. Rev. D}
  {\bfseries 106} no.~7, (2022) 076003},
  \href{http://arxiv.org/abs/2204.13119}{{\ttfamily arXiv:2204.13119
  [hep-th]}}.

\bibitem{Farina:2016rws}
M.~Farina, G.~Panico, D.~Pappadopulo, J.~T. Ruderman, R.~Torre, and A.~Wulzer,
  ``{Energy helps accuracy: electroweak precision tests at hadron colliders},''
  \href{http://dx.doi.org/10.1016/j.physletb.2017.06.043}{{\em Phys. Lett. B}
  {\bfseries 772} (2017) 210--215},
  \href{http://arxiv.org/abs/1609.08157}{{\ttfamily arXiv:1609.08157
  [hep-ph]}}.

\bibitem{Franceschini:2017xkh}
R.~Franceschini, G.~Panico, A.~Pomarol, F.~Riva, and A.~Wulzer, ``{Electroweak
  Precision Tests in High-Energy Diboson Processes},''
  \href{http://dx.doi.org/10.1007/JHEP02(2018)111}{{\em JHEP} {\bfseries 02}
  (2018) 111}, \href{http://arxiv.org/abs/1712.01310}{{\ttfamily
  arXiv:1712.01310 [hep-ph]}}.

\bibitem{Liu:2018pkg}
D.~Liu and L.-T. Wang, ``{Prospects for precision measurement of diboson
  processes in the semileptonic decay channel in future LHC runs},''
  \href{http://dx.doi.org/10.1103/PhysRevD.99.055001}{{\em Phys. Rev. D}
  {\bfseries 99} no.~5, (2019) 055001},
  \href{http://arxiv.org/abs/1804.08688}{{\ttfamily arXiv:1804.08688
  [hep-ph]}}.

\bibitem{Panico:2021vav}
G.~Panico, L.~Ricci, and A.~Wulzer, ``{High-energy EFT probes with fully
  differential Drell-Yan measurements},''
  \href{http://dx.doi.org/10.1007/JHEP07(2021)086}{{\em JHEP} {\bfseries 07}
  (2021) 086}, \href{http://arxiv.org/abs/2103.10532}{{\ttfamily
  arXiv:2103.10532 [hep-ph]}}.

\bibitem{Delahaye:2019omf}
J.~P. Delahaye, M.~Diemoz, K.~Long, B.~Mansouli\'e, N.~Pastrone, L.~Rivkin,
  D.~Schulte, A.~Skrinsky, and A.~Wulzer, ``{Muon Colliders},''
  \href{http://arxiv.org/abs/1901.06150}{{\ttfamily arXiv:1901.06150
  [physics.acc-ph]}}.

\bibitem{MuonCollider:2022nsa}
{\bfseries Muon Collider} Collaboration, D.~Stratakis {\em et~al.}, ``{A Muon
  Collider Facility for Physics Discovery},''
  \href{http://arxiv.org/abs/2203.08033}{{\ttfamily arXiv:2203.08033
  [physics.acc-ph]}}.

\bibitem{Chiesa:2020awd}
M.~Chiesa, F.~Maltoni, L.~Mantani, B.~Mele, F.~Piccinini, and X.~Zhao,
  ``{Measuring the quartic Higgs self-coupling at a multi-TeV muon collider},''
  \href{http://dx.doi.org/10.1007/JHEP09(2020)098}{{\em JHEP} {\bfseries 09}
  (2020) 098}, \href{http://arxiv.org/abs/2003.13628}{{\ttfamily
  arXiv:2003.13628 [hep-ph]}}.

\bibitem{Costantini:2020stv}
A.~Costantini, F.~De~Lillo, F.~Maltoni, L.~Mantani, O.~Mattelaer, R.~Ruiz, and
  X.~Zhao, ``{Vector boson fusion at multi-TeV muon colliders},''
  \href{http://dx.doi.org/10.1007/JHEP09(2020)080}{{\em JHEP} {\bfseries 09}
  (2020) 080}, \href{http://arxiv.org/abs/2005.10289}{{\ttfamily
  arXiv:2005.10289 [hep-ph]}}.

\bibitem{Capdevilla:2020qel}
R.~Capdevilla, D.~Curtin, Y.~Kahn, and G.~Krnjaic, ``{Discovering the physics
  of $(g-2)_\mu$ at future muon colliders},''
  \href{http://dx.doi.org/10.1103/PhysRevD.103.075028}{{\em Phys. Rev. D}
  {\bfseries 103} no.~7, (2021) 075028},
  \href{http://arxiv.org/abs/2006.16277}{{\ttfamily arXiv:2006.16277
  [hep-ph]}}.

\bibitem{Han:2020uid}
T.~Han, Y.~Ma, and K.~Xie, ``{High energy leptonic collisions and electroweak
  parton distribution functions},''
  \href{http://dx.doi.org/10.1103/PhysRevD.103.L031301}{{\em Phys. Rev. D}
  {\bfseries 103} no.~3, (2021) L031301},
  \href{http://arxiv.org/abs/2007.14300}{{\ttfamily arXiv:2007.14300
  [hep-ph]}}.

\bibitem{Han:2020pif}
T.~Han, D.~Liu, I.~Low, and X.~Wang, ``{Electroweak couplings of the Higgs
  boson at a multi-TeV muon collider},''
  \href{http://dx.doi.org/10.1103/PhysRevD.103.013002}{{\em Phys. Rev. D}
  {\bfseries 103} no.~1, (2021) 013002},
  \href{http://arxiv.org/abs/2008.12204}{{\ttfamily arXiv:2008.12204
  [hep-ph]}}.

\bibitem{Han:2020uak}
T.~Han, Z.~Liu, L.-T. Wang, and X.~Wang, ``{WIMPs at High Energy Muon
  Colliders},'' \href{http://dx.doi.org/10.1103/PhysRevD.103.075004}{{\em Phys.
  Rev. D} {\bfseries 103} no.~7, (2021) 075004},
  \href{http://arxiv.org/abs/2009.11287}{{\ttfamily arXiv:2009.11287
  [hep-ph]}}.

\bibitem{Buttazzo:2020ibd}
D.~Buttazzo and P.~Paradisi, ``{Probing the muon $g-2$ anomaly with the Higgs
  boson at a muon collider},''
  \href{http://dx.doi.org/10.1103/PhysRevD.104.075021}{{\em Phys. Rev. D}
  {\bfseries 104} no.~7, (2021) 075021},
  \href{http://arxiv.org/abs/2012.02769}{{\ttfamily arXiv:2012.02769
  [hep-ph]}}.

\bibitem{Yin:2020afe}
W.~Yin and M.~Yamaguchi, ``{Muon g-2 at a multi-TeV muon collider},''
  \href{http://dx.doi.org/10.1103/PhysRevD.106.033007}{{\em Phys. Rev. D}
  {\bfseries 106} no.~3, (2022) 033007},
  \href{http://arxiv.org/abs/2012.03928}{{\ttfamily arXiv:2012.03928
  [hep-ph]}}.

\bibitem{Buttazzo:2020uzc}
D.~Buttazzo, R.~Franceschini, and A.~Wulzer, ``{Two Paths Towards Precision at
  a Very High Energy Lepton Collider},''
  \href{http://dx.doi.org/10.1007/JHEP05(2021)219}{{\em JHEP} {\bfseries 05}
  (2021) 219}, \href{http://arxiv.org/abs/2012.11555}{{\ttfamily
  arXiv:2012.11555 [hep-ph]}}.

\bibitem{Huang:2021nkl}
G.-y. Huang, F.~S. Queiroz, and W.~Rodejohann, ``{Gauged
  $L^{}_{\mu}{-}L^{}_{\tau}$ at a muon collider},''
  \href{http://dx.doi.org/10.1103/PhysRevD.103.095005}{{\em Phys. Rev. D}
  {\bfseries 103} no.~9, (2021) 095005},
  \href{http://arxiv.org/abs/2101.04956}{{\ttfamily arXiv:2101.04956
  [hep-ph]}}.

\bibitem{Liu:2021jyc}
W.~Liu and K.-P. Xie, ``{Probing electroweak phase transition with multi-TeV
  muon colliders and gravitational waves},''
  \href{http://dx.doi.org/10.1007/JHEP04(2021)015}{{\em JHEP} {\bfseries 04}
  (2021) 015}, \href{http://arxiv.org/abs/2101.10469}{{\ttfamily
  arXiv:2101.10469 [hep-ph]}}.

\bibitem{Capdevilla:2021rwo}
R.~Capdevilla, D.~Curtin, Y.~Kahn, and G.~Krnjaic, ``{No-lose theorem for
  discovering the new physics of (g-2)\ensuremath{\mu} at muon colliders},''
  \href{http://dx.doi.org/10.1103/PhysRevD.105.015028}{{\em Phys. Rev. D}
  {\bfseries 105} no.~1, (2022) 015028},
  \href{http://arxiv.org/abs/2101.10334}{{\ttfamily arXiv:2101.10334
  [hep-ph]}}.

\bibitem{Han:2021udl}
T.~Han, S.~Li, S.~Su, W.~Su, and Y.~Wu, ``{Heavy Higgs bosons in 2HDM at a muon
  collider},'' \href{http://dx.doi.org/10.1103/PhysRevD.104.055029}{{\em Phys.
  Rev. D} {\bfseries 104} no.~5, (2021) 055029},
  \href{http://arxiv.org/abs/2102.08386}{{\ttfamily arXiv:2102.08386
  [hep-ph]}}.

\bibitem{Capdevilla:2021fmj}
R.~Capdevilla, F.~Meloni, R.~Simoniello, and J.~Zurita, ``{Hunting wino and
  higgsino dark matter at the muon collider with disappearing tracks},''
  \href{http://dx.doi.org/10.1007/JHEP06(2021)133}{{\em JHEP} {\bfseries 06}
  (2021) 133}, \href{http://arxiv.org/abs/2102.11292}{{\ttfamily
  arXiv:2102.11292 [hep-ph]}}.

\bibitem{Han:2021kes}
T.~Han, Y.~Ma, and K.~Xie, ``{Quark and gluon contents of a lepton at high
  energies},'' \href{http://dx.doi.org/10.1007/JHEP02(2022)154}{{\em JHEP}
  {\bfseries 02} (2022) 154}, \href{http://arxiv.org/abs/2103.09844}{{\ttfamily
  arXiv:2103.09844 [hep-ph]}}.

\bibitem{AlAli:2021let}
H.~Al~Ali {\em et~al.}, ``{The muon Smasher\textquoteright{}s guide},''
  \href{http://dx.doi.org/10.1088/1361-6633/ac6678}{{\em Rept. Prog. Phys.}
  {\bfseries 85} no.~8, (2022) 084201},
  \href{http://arxiv.org/abs/2103.14043}{{\ttfamily arXiv:2103.14043
  [hep-ph]}}.

\bibitem{Asadi:2021gah}
P.~Asadi, R.~Capdevilla, C.~Cesarotti, and S.~Homiller, ``{Searching for
  leptoquarks at future muon colliders},''
  \href{http://dx.doi.org/10.1007/JHEP10(2021)182}{{\em JHEP} {\bfseries 10}
  (2021) 182}, \href{http://arxiv.org/abs/2104.05720}{{\ttfamily
  arXiv:2104.05720 [hep-ph]}}.

\bibitem{Bottaro:2021snn}
S.~Bottaro, D.~Buttazzo, M.~Costa, R.~Franceschini, P.~Panci, D.~Redigolo, and
  L.~Vittorio, ``{Closing the window on WIMP Dark Matter},''
  \href{http://dx.doi.org/10.1140/epjc/s10052-021-09917-9}{{\em Eur. Phys. J.
  C} {\bfseries 82} no.~1, (2022) 31},
  \href{http://arxiv.org/abs/2107.09688}{{\ttfamily arXiv:2107.09688
  [hep-ph]}}.

\bibitem{Qian:2021ihf}
S.~Qian, C.~Li, Q.~Li, F.~Meng, J.~Xiao, T.~Yang, M.~Lu, and Z.~You,
  ``{Searching for heavy leptoquarks at a muon collider},''
  \href{http://dx.doi.org/10.1007/JHEP12(2021)047}{{\em JHEP} {\bfseries 12}
  (2021) 047}, \href{http://arxiv.org/abs/2109.01265}{{\ttfamily
  arXiv:2109.01265 [hep-ph]}}.

\bibitem{Chiesa:2021qpr}
M.~Chiesa, B.~Mele, and F.~Piccinini, ``{Multi Higgs production via photon
  fusion at future multi-TeV muon colliders},''
  \href{http://arxiv.org/abs/2109.10109}{{\ttfamily arXiv:2109.10109
  [hep-ph]}}.

\bibitem{Liu:2021akf}
W.~Liu, K.-P. Xie, and Z.~Yi, ``{Testing leptogenesis at the LHC and future
  muon colliders: A Z' scenario},''
  \href{http://dx.doi.org/10.1103/PhysRevD.105.095034}{{\em Phys. Rev. D}
  {\bfseries 105} no.~9, (2022) 095034},
  \href{http://arxiv.org/abs/2109.15087}{{\ttfamily arXiv:2109.15087
  [hep-ph]}}.

\bibitem{Chen:2021pqi}
J.~Chen, T.~Li, C.-T. Lu, Y.~Wu, and C.-Y. Yao, ``{Measurement of Higgs boson
  self-couplings through 2\textrightarrow{}3 vector bosons scattering in future
  muon colliders},'' \href{http://dx.doi.org/10.1103/PhysRevD.105.053009}{{\em
  Phys. Rev. D} {\bfseries 105} no.~5, (2022) 053009},
  \href{http://arxiv.org/abs/2112.12507}{{\ttfamily arXiv:2112.12507
  [hep-ph]}}.

\bibitem{Chen:2022msz}
S.~Chen, A.~Glioti, R.~Rattazzi, L.~Ricci, and A.~Wulzer, ``{Learning from
  Radiation at a Very High Energy Lepton Collider},''
  \href{http://arxiv.org/abs/2202.10509}{{\ttfamily arXiv:2202.10509
  [hep-ph]}}.

\bibitem{Cesarotti:2022ttv}
C.~Cesarotti, S.~Homiller, R.~K. Mishra, and M.~Reece, ``{Probing New Gauge
  Forces with a High-Energy Muon Beam Dump},''
  \href{http://arxiv.org/abs/2202.12302}{{\ttfamily arXiv:2202.12302
  [hep-ph]}}.

\bibitem{deBlas:2022aow}
J.~de~Blas, J.~Gu, and Z.~Liu, ``{Higgs boson precision measurements at a
  125~GeV muon collider},''
  \href{http://dx.doi.org/10.1103/PhysRevD.106.073007}{{\em Phys. Rev. D}
  {\bfseries 106} no.~7, (2022) 073007},
  \href{http://arxiv.org/abs/2203.04324}{{\ttfamily arXiv:2203.04324
  [hep-ph]}}.

\bibitem{Bao:2022onq}
Y.~Bao, J.~Fan, and L.~Li, ``{Electroweak ALP searches at a muon collider},''
  \href{http://dx.doi.org/10.1007/JHEP08(2022)276}{{\em JHEP} {\bfseries 08}
  (2022) 276}, \href{http://arxiv.org/abs/2203.04328}{{\ttfamily
  arXiv:2203.04328 [hep-ph]}}.

\bibitem{Homiller:2022iax}
S.~Homiller, Q.~Lu, and M.~Reece, ``{Complementary signals of lepton flavor
  violation at a high-energy muon collider},''
  \href{http://dx.doi.org/10.1007/JHEP07(2022)036}{{\em JHEP} {\bfseries 07}
  (2022) 036}, \href{http://arxiv.org/abs/2203.08825}{{\ttfamily
  arXiv:2203.08825 [hep-ph]}}.

\bibitem{Forslund:2022xjq}
M.~Forslund and P.~Meade, ``{High precision higgs from high energy muon
  colliders},'' \href{http://dx.doi.org/10.1007/JHEP08(2022)185}{{\em JHEP}
  {\bfseries 08} (2022) 185}, \href{http://arxiv.org/abs/2203.09425}{{\ttfamily
  arXiv:2203.09425 [hep-ph]}}.

\bibitem{Csaki:2008zd}
C.~Csaki, A.~Falkowski, and A.~Weiler, ``{The Flavor of the Composite
  Pseudo-Goldstone Higgs},''
  \href{http://dx.doi.org/10.1088/1126-6708/2008/09/008}{{\em JHEP} {\bfseries
  09} (2008) 008}, \href{http://arxiv.org/abs/0804.1954}{{\ttfamily
  arXiv:0804.1954 [hep-ph]}}.

\bibitem{Keren-Zur:2012buf}
B.~Keren-Zur, P.~Lodone, M.~Nardecchia, D.~Pappadopulo, R.~Rattazzi, and
  L.~Vecchi, ``{On Partial Compositeness and the CP asymmetry in charm
  decays},'' \href{http://dx.doi.org/10.1016/j.nuclphysb.2012.10.012}{{\em
  Nucl. Phys. B} {\bfseries 867} (2013) 394--428},
  \href{http://arxiv.org/abs/1205.5803}{{\ttfamily arXiv:1205.5803 [hep-ph]}}.

\bibitem{Dawson:1984gx}
S.~Dawson, ``{The Effective W Approximation},''
  \href{http://dx.doi.org/10.1016/0550-3213(85)90038-0}{{\em Nucl. Phys. B}
  {\bfseries 249} (1985) 42--60}.

\bibitem{Kunszt:1987tk}
Z.~Kunszt and D.~E. Soper, ``{On the Validity of the Effective $W$
  Approximation},'' \href{http://dx.doi.org/10.1016/0550-3213(88)90673-6}{{\em
  Nucl. Phys. B} {\bfseries 296} (1988) 253--289}.

\bibitem{Borel:2012by}
P.~Borel, R.~Franceschini, R.~Rattazzi, and A.~Wulzer, ``{Probing the
  Scattering of Equivalent Electroweak Bosons},''
  \href{http://dx.doi.org/10.1007/JHEP06(2012)122}{{\em JHEP} {\bfseries 06}
  (2012) 122}, \href{http://arxiv.org/abs/1202.1904}{{\ttfamily arXiv:1202.1904
  [hep-ph]}}.

\bibitem{Azatov:2017kzw}
A.~Azatov, J.~Elias-Miro, Y.~Reyimuaji, and E.~Venturini, ``{Novel measurements
  of anomalous triple gauge couplings for the LHC},''
  \href{http://dx.doi.org/10.1007/JHEP10(2017)027}{{\em JHEP} {\bfseries 10}
  (2017) 027}, \href{http://arxiv.org/abs/1707.08060}{{\ttfamily
  arXiv:1707.08060 [hep-ph]}}.

\bibitem{Panico:2017frx}
G.~Panico, F.~Riva, and A.~Wulzer, ``{Diboson interference resurrection},''
  \href{http://dx.doi.org/10.1016/j.physletb.2017.11.068}{{\em Phys. Lett. B}
  {\bfseries 776} (2018) 473--480},
  \href{http://arxiv.org/abs/1708.07823}{{\ttfamily arXiv:1708.07823
  [hep-ph]}}.

\bibitem{Maltoni:2019}
F.~Maltoni, L.~Mantani, and K.~Mimasu, ``{Top-quark electroweak interactions at
  high energy},'' \href{http://dx.doi.org/10.1007/JHEP10(2019)004}{{\em JHEP}
  {\bfseries 10} (2019) 004}, \href{http://arxiv.org/abs/1904.05637}{{\ttfamily
  arXiv:1904.05637 [hep-ph]}}.

\bibitem{LHCHiggsCrossSectionWorkingGroup:2013rie}
{\bfseries LHC Higgs Cross Section Working Group} Collaboration, J.~R. Andersen
  {\em et~al.}, ``{Handbook of LHC Higgs Cross Sections: 3. Higgs
  Properties},'' \href{http://arxiv.org/abs/1307.1347}{{\ttfamily
  arXiv:1307.1347 [hep-ph]}}.

\bibitem{CLICdp:2018esa}
{\bfseries CLICdp} Collaboration, H.~Abramowicz {\em et~al.}, ``{Top-Quark
  Physics at the CLIC Electron-Positron Linear Collider},''
  \href{http://dx.doi.org/10.1007/JHEP11(2019)003}{{\em JHEP} {\bfseries 11}
  (2019) 003}, \href{http://arxiv.org/abs/1807.02441}{{\ttfamily
  arXiv:1807.02441 [hep-ex]}}.

\bibitem{madgraph}
J.~Alwall, R.~Frederix, S.~Frixione, V.~Hirschi, F.~Maltoni, O.~Mattelaer,
  H.~S. Shao, T.~Stelzer, P.~Torrielli, and M.~Zaro, ``{The automated
  computation of tree-level and next-to-leading order differential cross
  sections, and their matching to parton shower simulations},''
  \href{http://dx.doi.org/10.1007/JHEP07(2014)079}{{\em JHEP} {\bfseries 07}
  (2014) 079}, \href{http://arxiv.org/abs/1405.0301}{{\ttfamily arXiv:1405.0301
  [hep-ph]}}.

\bibitem{bsmc}
A.~Falkowski, B.~Fuks, K.~Mawatari, K.~Mimasu, F.~Riva, and V.~Sanz,
  ``{Rosetta: an operator basis translator for Standard Model effective field
  theory},'' \href{http://dx.doi.org/10.1140/epjc/s10052-015-3806-x}{{\em Eur.
  Phys. J. C} {\bfseries 75} no.~12, (2015) 583},
  \href{http://arxiv.org/abs/1508.05895}{{\ttfamily arXiv:1508.05895
  [hep-ph]}}.

\bibitem{ParticleDataGroup:2020ssz}
{\bfseries Particle Data Group} Collaboration, P.~A. Zyla {\em et~al.},
  ``{Review of Particle Physics},''
  \href{http://dx.doi.org/10.1093/ptep/ptaa104}{{\em PTEP} {\bfseries 2020}
  no.~8, (2020) 083C01}.

\bibitem{Workman:2022ynf}
{\bfseries Particle Data Group} Collaboration, R.~L. Workman and Others,
  ``{Review of Particle Physics},''
  \href{http://dx.doi.org/10.1093/ptep/ptac097}{{\em PTEP} {\bfseries 2022}
  (2022) 083C01}.

\bibitem{Cepeda:2019klc}
M.~Cepeda {\em et~al.}, ``{Report from Working Group 2}: {Higgs Physics at the
  HL-LHC and HE-LHC},''
  \href{http://dx.doi.org/10.23731/CYRM-2019-007.221}{{\em CERN Yellow Rep.
  Monogr.} {\bfseries 7} (2019) 221--584},
  \href{http://arxiv.org/abs/1902.00134}{{\ttfamily arXiv:1902.00134
  [hep-ph]}}.

\bibitem{Mangano:2015aow}
M.~L. Mangano, T.~Plehn, P.~Reimitz, T.~Schell, and H.-S. Shao, ``{Measuring
  the Top Yukawa Coupling at 100 TeV},''
  \href{http://dx.doi.org/10.1088/0954-3899/43/3/035001}{{\em J. Phys. G}
  {\bfseries 43} no.~3, (2016) 035001},
  \href{http://arxiv.org/abs/1507.08169}{{\ttfamily arXiv:1507.08169
  [hep-ph]}}.

\bibitem{Alwall:2014hca}
J.~Alwall, R.~Frederix, S.~Frixione, V.~Hirschi, F.~Maltoni, O.~Mattelaer,
  H.~S. Shao, T.~Stelzer, P.~Torrielli, and M.~Zaro, ``{The automated
  computation of tree-level and next-to-leading order differential cross
  sections, and their matching to parton shower simulations},''
  \href{http://dx.doi.org/10.1007/JHEP07(2014)079}{{\em JHEP} {\bfseries 07}
  (2014) 079}, \href{http://arxiv.org/abs/1405.0301}{{\ttfamily arXiv:1405.0301
  [hep-ph]}}.

\bibitem{Hagiwara:1986vm}
K.~Hagiwara, R.~D. Peccei, D.~Zeppenfeld, and K.~Hikasa, ``{Probing the Weak
  Boson Sector in e+ e- ---\ensuremath{>} W+ W-},''
  \href{http://dx.doi.org/10.1016/0550-3213(87)90685-7}{{\em Nucl. Phys. B}
  {\bfseries 282} (1987) 253--307}.

\bibitem{Giudice:2007fh}
G.~F. Giudice, C.~Grojean, A.~Pomarol, and R.~Rattazzi, ``{The
  Strongly-Interacting Light Higgs},''
  \href{http://dx.doi.org/10.1088/1126-6708/2007/06/045}{{\em JHEP} {\bfseries
  06} (2007) 045}, \href{http://arxiv.org/abs/hep-ph/0703164}{{\ttfamily
  arXiv:hep-ph/0703164}}.

\end{thebibliography}\endgroup

\end{document}